\documentclass[9pt,twocolumn,twoside
]{pnas-new}

\templatetype{pnasresearcharticle} 
\setboolean{displaywatermark}{false}

\DeclareUnicodeCharacter{2061}{}

\usepackage{lipsum} 

\usepackage{eqnarray}

\usepackage{graphicx}
\usepackage{dcolumn}
\DeclareUnicodeCharacter{2061}{}
\usepackage{bm}
\usepackage{bbold}
\usepackage{mathtools}

\newcommand\asm[1]{}

\newcommand{\nyuphysics}{Center for Soft Matter Research, Department of Physics, New York University, New York 10003, USA}

\newcommand{\nyusimons}{Simons Center for Computational Physical Chemistry, Department of Chemistry, New York University, New York 10003, USA}
\newcommand{\nyucourant}{Courant Institute of Mathematical Sciences, New York University, New York 10003, USA}

\newcommand{\tauphys}{School of Physics and Astronomy, and the Center for Physics and Chemistry of Living Systems, Tel Aviv University, Tel Aviv 6997801, Israel}

\newcommand{\radboud}{Department of Artificial Intelligence, Donders Center for Cognition, Radboud University, Nijmegen, Netherlands}

\usepackage{setspace}

\newcolumntype{C}[1]{>{\centering}m{#1}}

\begin{document}

\title{A geometric condition for robot-swarm cohesion and cluster-flock transition}

\author[a,b]{Mathias Casiulis}
\author[c]{Eden Arbel}
\author[d]{Charlotte van Waes}
\author[c]{Yoav Lahini}
\author[a,b,e]{Stefano Martiniani}
\author[c]{Naomi Oppenheimer}
\author[d,1]{Matan Yah Ben Zion}

\affil[a]{\nyuphysics}
\affil[b]{\nyusimons}
\affil[c]{\tauphys}
\affil[d]{\radboud}
\affil[e]{\nyucourant}

\leadauthor{Casiulis} 

\significancestatement{
Robotic swarms, ensembles of collaborative robots that work together to achieve tasks, are an appealing solution to tackle complex tasks such as automated exploration, foraging, or transport.
Yet, a scalable swarm cannot rely on an external controller nor complex computation, and requires simple design rules to achieve emergent functions.
Viewing robots as self-propelled particles, we show that the size and mass repartition of an individual robot define an intrinsic curvature.
This curvature seeds the collective behavior of the swarm, offering a direct design rule to control whether the swarm flocks, flows, or clusters.
We thus demonstrate a computation-free route for decentralized control on collective behavior, paving the way for richer swarm robotic applications.
}

\authorcontributions{
M. C. contributed to the design and implementation of the simulations, the data generation and analysis, and writing of this manuscript.
E. A. contributed to experimental design, execution, and data analysis.
S. M. contributed to the conceptualization,  data analysis, writing of this manuscript, and funding acquisition.
C.v.W Contributed to the derivation of the analytical model.
Y.L. Contributed to the experimental design.
N.O. Contributed to the analytical model and the development of the numerical engine.
M.Y.B.Z. Initiated the project and supervised its execution, designed its experimental,  numerical, and analytical components, wrote the first draft of the manuscript, and contributed to funding acquisition.}
\authordeclaration{The authors have no conflict of interest to declare.}
\correspondingauthor{\textsuperscript{1}E-mail: matanbz@gmail.com}

\keywords{ Robophysics $|$ Swarm Robotics $|$ Active Matter $|$ Clustering $|$ Flocking $|$ Decentralized Control}

\begin{abstract}
We present a geometric design rule for size-controlled clustering of self-propelled particles.
We show that active particles that tend to rotate under an external force have an intrinsic, signed parameter with units of curvature which we call curvity, that can be derived from first principles.
Experiments with robots and numerical simulations show that properties of individual robots (radius and curvity) control pair cohesion in a binary system, and the stability of flocking and self-limiting clustering in a swarm, with applications in meta-materials and in embodied decentralized control.
\end{abstract}

\dates{This manuscript was compiled on \today}
\doi{\url{www.pnas.org/cgi/doi/10.1073/pnas.XXXXXXXXXX}}

\maketitle
\newpage 

\thispagestyle{firststyle}
\ifthenelse{\boolean{shortarticle}}{\ifthenelse{\boolean{singlecolumn}}{\abscontentformatted}{\abscontent}}{}

\dropcap{A}ctive matter offers a wealth of behaviors unfathomable at equilibrium: the broken time-reversal-symmetry and lack of Galilean invariance~\cite{Toner2005,Nardini2017,Modin2023} unlock new dynamical~\cite{Vicsek1995,Casiulis2020,Fruchart2021}, structural~\cite{Lei2019,Casiulis2019b,Levis2019PRE,Zhang2022, Casiulis2022a,Shi2023,Adorjáni2024,Lei2024}, and functional~\cite{Levis2019PRR,Ro2022,BenZion2022,Anand2024} states, that expand the notion of materials to describe both living systems and robotic swarms~\cite{Tajima2002,Cavagna2014,Rubinstein2014, Alonso-Marroquin2014,Cavagna2018,Sugi2019,Wang2021,Boudet2021}.
At equilibrium, the direct link between pair interactions and the emergent structures is established by statistical mechanics~\cite{Weeks1971,Hansen2006,Kardar2007}, with applications for multi-scale design~\cite{whiteside2002,Glotzer2004,Seeman2015, Frenkel2015,BenZion2017,Hueckel2021a}.
Far from equilibrium however this luxury is largely absent.

Save for a handful of systems~\cite{Oppenheimer2022,Shoham2023}, the only solution is often hydrodynamic-level theories~\cite{Onsager1949,Bertin2009,Peshkov2014,NguyenThuLam2015,NguyenThuLam2015a,Chate2020}, that rely on coarse-graining interactions in the dilute limit.
In systems of self-propelled particles, such approaches have proven precious to predict the emergence of collective behaviors like flocking, when a particle aligns with the average heading of its neighbors~\cite{Vicsek1995,Marchetti2013}, Motility-Induced Phase Separation (MIPS)~\cite{Tailleur2008,Cates2014,Solon2018}, observed when a particle slows down with increasing local particle concentration, or a similar crowding observed when a particle's heading turns towards locally higher particle concentrations~\cite{Zhang2021b}.
Yet, it remains arduous to design functional active materials bottom-up --- predicting the collective response of a large ensemble first requires knowledge of the response of an individual to a smaller ensemble.

Microscopic models of self-propelled particles typically describe the direct response of their velocity, $\vec{v}$, to external forces, $\vec{F}$, through an effective mobility, $\mu$~\cite{Winkler2015,Caprini2020,Caprini2021b}.
It was identified empirically that 
forces also generically couple to the particle's orientation, $\hat{e}$, effectively imposing a torque that rotates its heading~\cite{Shimoyama1996}.
Thus, many works considered such torques in effective models~\cite{Weber2013, Dauchot2019,Mirhosseini2022,Baconnier2022,Baconnier2024}.
Yet, these works lacked a microscopic mechanical model for them, so that they were only assumed to be proportional to either $\vec{v}$~\cite{Baconnier2022} or $\hat{v}$~\cite{Shimoyama1996,Weber2013,Dauchot2019}, and that their prefactor was assumed to be non-negative ($\hat{e}$ aligns onto $\vec{F}$) because it was identified to a persistence time or length.
Recently, a microscopic model was proposed for the dynamics of hopping self-propelled particles~\cite{Arbel2024}, revealing in particular torques proportional to $\vec{v}$, and that the prefactor of the torque should be seen as an activity-induced quantity with units of curvature, thus dubbed \textit{curvity}.
Crucially, curvity is signed, much like an electric charge.
When positive ($\hat{e}$ aligns onto $\vec{F}$), it leads to orbiting dynamics in a confining potential ~\cite{Dauchot2019}, and to flocking through collision induced alignment in ensembles~\cite{Weber2013,NguyenThuLam2015,NguyenThuLam2015a,Das2024}.
When negative  ($\hat{e}$ anti-aligns with $\vec{F}$), it couples to the curvature of passive objects, which can induce cooperative transport of a movable payload through spontaneous symmetry breaking~\cite{Arbel2024}.

In this paper, we show experimentally, numerically, and analytically, that the \textit{curvity} also couples to the curvature of the self-propelled particles themselves.
Pairs of particles of radii $b$, and curvity $\kappa$, will display effective attraction when the geometric criterion
\begin{equation}
    \kappa + 1/b < 0
    \label{eqAttraction}
\end{equation}
is met, which is the main result of our work.
Inequality~\ref{eqAttraction} is geometric in the sense that it compares two intrinsic length scales of the active particle, its curvity, $\kappa$, and its curvature, $1/b$ (inverse radius), and depends neither on kinematics (speeds and rates) nor mechanics (forces).
Note that the curvity does not depend on self-propulsion. An active particle can have a finite curvity ($\kappa \neq 0$) and rotate by an external force, even at the absence of self-propulsion ($v_0\rightarrow 0$, see SI for further details).
It conditions the attraction of an active particle to {\it another} active particle, or pair \textit{cohesion} and thus extends previous result on the \textit{adhesion} of a single active particle to a static potential~\cite{Arbel2024}.
We find this criterion holds in experiments using custom-built vibration-driven robots, as well as in Langevin simulations of self-propelled particles, where we varied  the particles diameters and curvities.
We also show through numerical simulations that Inequality~\ref{eqAttraction} sets the cornerstone for the many-body behavior.
We present an extension of this criterion to finite densities, that predicts a clustering transition into crystallites of controlled size, paving the way for controlling the large scale behavior of self-propelled particles such as robotic swarms using only embodied parameters.

\begin{figure}[t]
    \centering
    \includegraphics[width=1\linewidth]{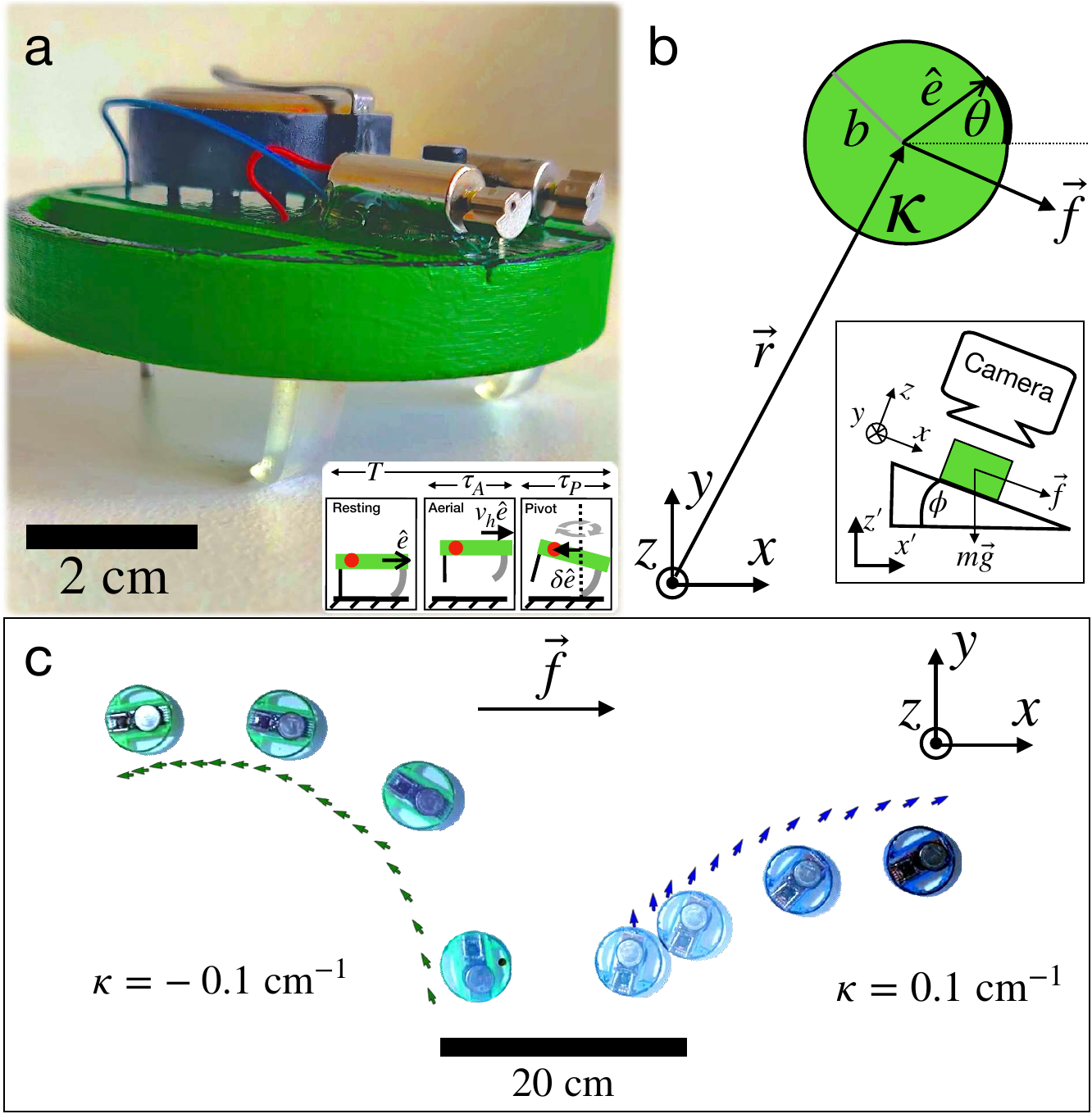}
    \caption{Measuring the curvity, $\kappa$, of a self-propelled particle. 
    (a) A vibration-driven robot showing two vibration motors, two soft front legs, and a stiff back leg. Inset illustrates the displaced ($\vec{\delta}$) center of mass (red dot) from the rotation axis (dashed line) along the heading ($\hat{e}$) in the aerial and pivot phases of motion resulting in Eqs.~\ref{eqForce},~\ref{eqAlignment}.
    (b) The curvity, $\kappa$, is measured using an inclined plane to apply a constant force (inset), monitoring the particle's position ($\vec{r}$) and heading ($\hat{e}$).
    (c) Heading rotation along (blue $\kappa >0$) or against (green $\kappa < 0$) an external constant force (see vidoes in SM).
    }
    \label{fig01}
    \vspace{-0.5cm}
\end{figure}

We start by analytically investigating the deterministic pair dynamics, showing that the condition in Eq.~\ref{eqAttraction} offers a stable fixed point for pairwise attraction.
We then test this criterion experimentally by measuring the \textit{kissing time} ($\tau_k$) --- the duration over which two vibration-driven robots osculate.
Finally, we test this condition numerically by simulating the many-body dynamics over a range of densities, finding that the condition in Eq.~\ref{eqAttraction} naturally extends to finite concentrations and noise, and quantitatively predicts the size of self-limiting clusters.

{\it Equations of motion}. 
Empirically, the time evolution of the velocity, $\vec{v}$, and heading $\hat{e}\equiv\left(\cos \theta,\sin\theta\right)$, of a self-propelled particle placed in an external force, $\vec{f}$, is often captured~\cite{Chate2020,Baconnier2024} by the effective equations of motion
\begin{align}
    \frac{d}{dt}\vec{r}(t) &\equiv \vec{v}(t) = v_0 \hat{e}(t) + \mu \vec{f}(t),
    \label{eqForce}\\
    \frac{d}{dt} \hat{e}(t) &= \kappa \hat{e}(t) \times \left(\vec{v}(t)\times\hat{e}(t)\right),
    \label{eqAlignment}
\end{align}
where $v_0$ is the nominal speed, $\mu$ the mobility, and $\kappa$ the curvity of the particle.
The curvity, $\kappa$, is the first $\kappa$ey parameter in our geometric construction.
When an external force is perpendicular to the orientation of the self-propelled particle, the curvity determines how much the trajectory curves to align parallel ($\kappa > 0$) or anti-parallel ($\kappa <0$) to the force.
The following analogies from electrodynamics and hydrodynamics may be useful in appreciating the role of curvity: similar to the cyclotron orbit of an electric charge moving through an out of plane magnetic field ($R_L \propto 1/q$, where $q$ is the charge)~\cite{Jackson}, and the lift caused by the Magnus effect forcing a rotating disc moving through an ideal fluid into a curved trajectory ($R_M\propto 1/\Gamma$, where $\Gamma$ is the circulation)~\cite{Childress2009a}, a self-propelled particle subjected to a force field perpendicular to its heading will orbit with a radius $R \propto 1/\kappa$ (where $\kappa$ is its curvity)~\cite{BenZion2023}.

Equations~\ref{eqForce} and ~\ref{eqAlignment} were recently derived from a Newtonian description of vibration-driven robots~\cite{Arbel2024}, offering their microscopic origin, and a handle for swarm design.
In particular the curvity is $\kappa\equiv \delta \left(\tau_P/\tau_A\right)^2 m/I $, where $\tau_A$ and $\tau_P$ are the aloft and pivot times; $m$ and $I$ are the mass and moment of inertia; and $\delta$ is the signed displacement of the center of mass from the pivot axis along the heading, $\vec{\delta} \equiv \delta \hat{e}$ (see Fig.~\ref{fig01} and SM~\cite{supp}). 
Beyond this example, force-alignment can be pivotal in a large class of ``dry'' active matter, from shaken granules to bacterial colonies~\cite{Marchetti2013,Anand2024} --- except for the singular case where the center of mass is directly aligned with the center of stress, one should expect a finite (and signed) curvity.

\begin{figure*}[t]
    \centering
    \includegraphics[width=1\linewidth]{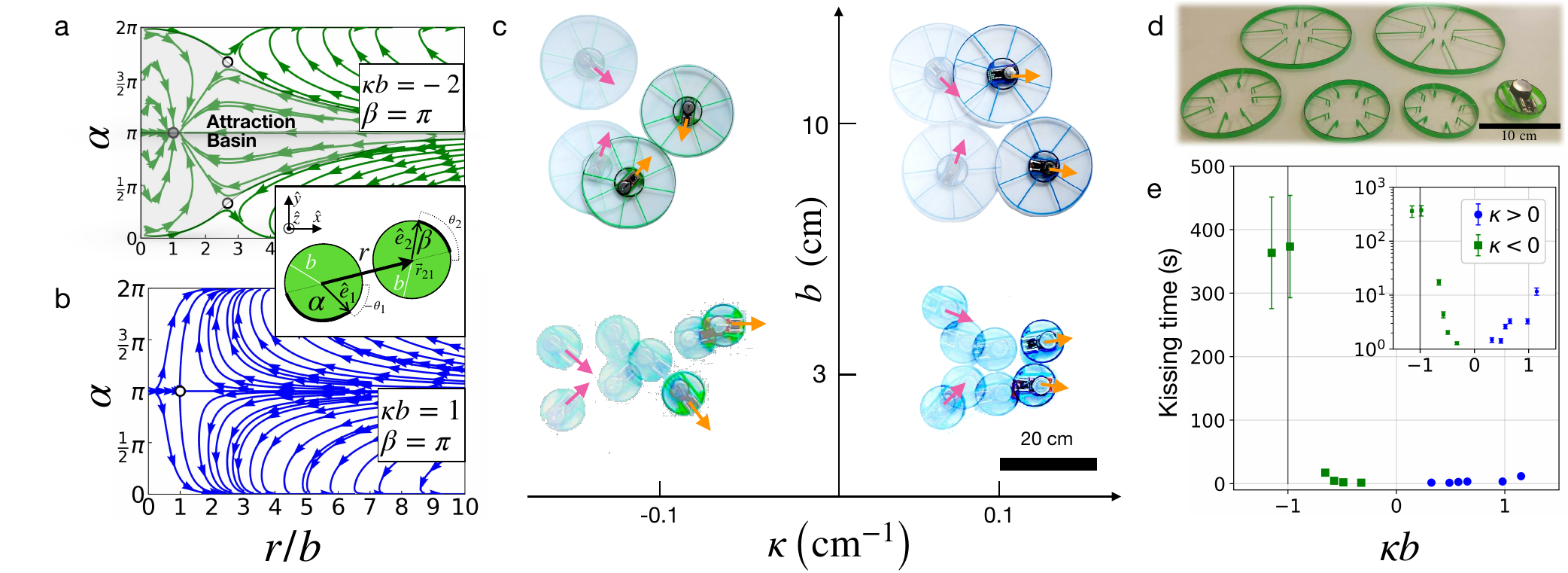}
    \caption{A geometric criterion for effective pairwise attraction of two self-propelled particles. {\bf (a)} and {\bf (b)}  and show $\alpha$-$r$ phase portrait of Eqs.~\ref{eqPairR},~\ref{eqPairAlpha} at a constant $\beta = \pi$. An attraction basin is found when the criterion for effective attraction (Eq.~\ref{eqAttraction}) is satisfied. Inset shows the dynamic variables of two interacting self-propelled particles. {\bf (c)} The initial (pink) and final (orange) states of pair-collision in vibration driven-robots show effective binding by fronting ($\alpha=\beta\rightarrow \pi$, $r\rightarrow 2b$) when Eq.~\ref{eqAttraction} is satisfied (top left), otherwise the robots scatter. {\bf (d)} 3D printed skirts of different sizes to change a robot's radius, $b$. {\bf (e)} Experimentally measured kissing time ($\tau_k$) shows over 2 orders of magnitude increase when $\kappa b < -1$, inline with the geometric criterion of attraction (Eq.~\ref{eqAttraction}, inset show log-scale.).}
    \label{fig02}
    \vspace{-0.5cm}
\end{figure*}

\begin{figure}[bh!]
    \centering
    \includegraphics[width=\linewidth]{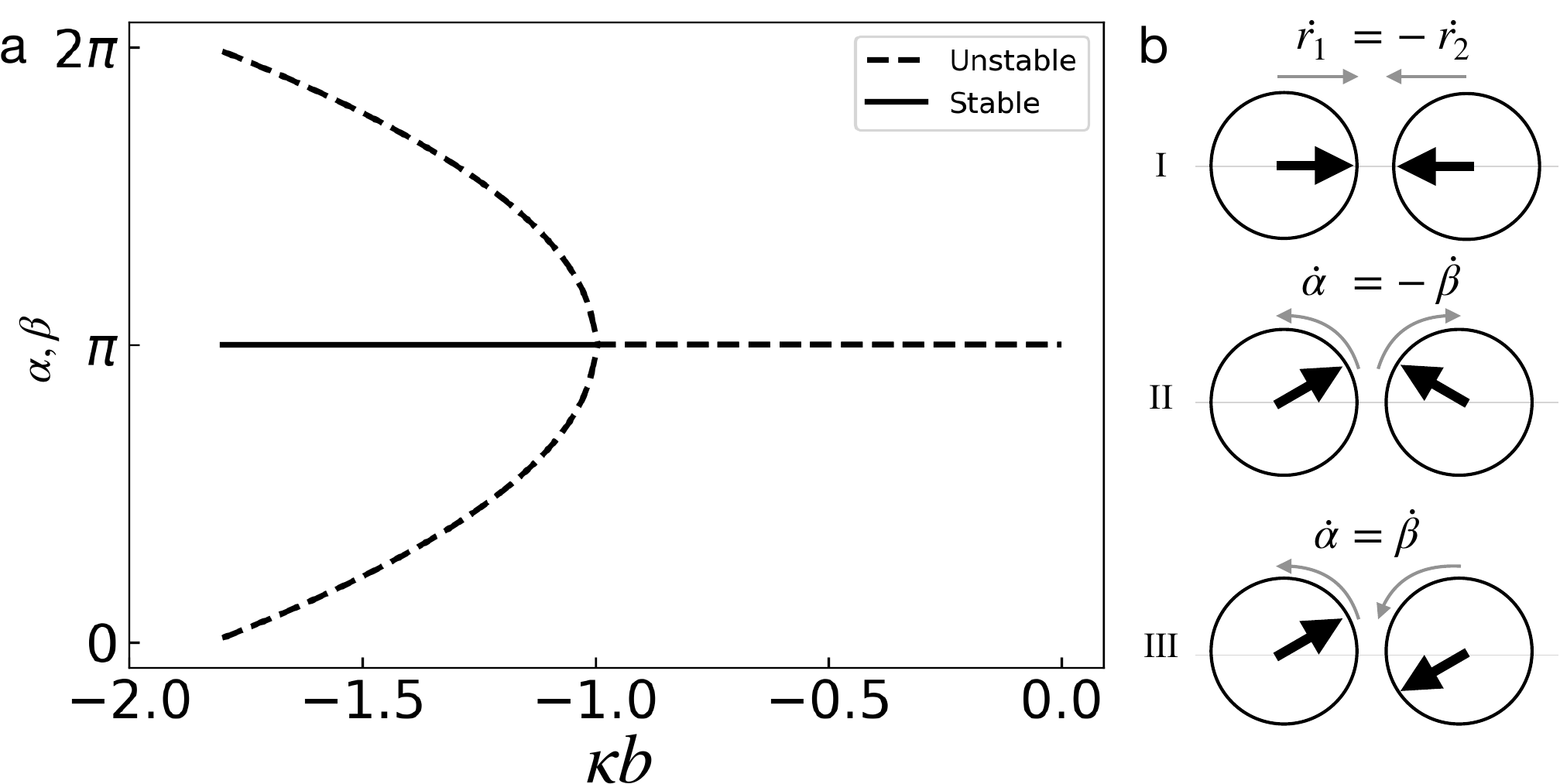}
    \caption{\textbf{Normal form around the stable equilibrium point.}
    (a) The pair attraction shows a sub-critical pitchfork bifurcation at $\kappa b = -1$.
    (b) The normal modes at the bifurcation are 
    (I)  motion along particles' centers ($\dot r_1 =-\dot r_2$),
    (II) Counter rotation ($\dot \alpha = -\dot \beta$),
    (III) co-rotation ($\dot \alpha = \dot \beta$). The first two (I and II) are stable even for $\kappa b >-1$, and the latter (III) is restoring only when $\kappa b <-1$.
    }
    \label{fig03}
    \vspace{-0.5cm}
\end{figure}

The second key parameter in the geometric construction is simply the radius of each particle, $b$.
When the steric repulsion between circular particles 1 and 2 is described as a radially symmetric force field, $\vec{F}_{21}$, their mutual dynamics follow
\begin{align}
    \vec{v}_{2} &= v_0 \hat{e}_2 + \mu  \vec{F}_{21}
    \label{eqForce21}\\
    \frac{d}{dt} \hat{e}_2 &= \kappa \hat{e}_2 \times \left(\vec{v}_2\times\hat{e}_2\right),
    \label{eqAlignment22}
\end{align}
and $2\rightarrow 1$ for particle 1 (see Fig.~\ref{fig02}a inset).
The point where steric repulsion balances self-propulsion, $\Gamma\left(r_{21} = 2b\right) \equiv 1$, defines the particles' radii.
When the repulsion is spatially decaying ($\Gamma ' <0$) the force profile $\Gamma$ can be left implicit, and apply for common steric interactions (soft-core, screened Coulomb, WCA etc.~\cite{Weeks1971,Thompson2022}), making the following argument general.
The 6 degrees of freedom in Eqs.~\ref{eqForce21},~\ref{eqAlignment22} reduce to only 3 relative degrees of freedom in center-of-mass coordinates: the distance between the centers, $|\vec{r}_{21}|=r$, and the relative angle of the headings $\alpha$, $\beta$
\begin{align}
    \dot r &= \;\;v_0 \left[\cos \alpha + \cos \beta + 2\Gamma\left(r\right)\right]\label{eqPairR}\\
    \dot\alpha &= -v_0 \left[\left(\kappa  \Gamma \left(r\right) + \frac{1}{r} \right) \sin \alpha + \frac{1}{r}\sin \beta\right]
    \label{eqPairAlpha}\\
    \dot\beta &= -v_0 \left[\left(\kappa  \Gamma \left(r\right) + \frac{1}{r} \right) \sin \beta + \frac{1}{r}\sin \alpha\right],
    \label{eqPairBeta}
\end{align}
see Fig.~\ref{fig02}a inset and SM~\cite{supp}. 
When the condition in Eq.~\ref{eqAttraction} is met, the dynamical system in Eqs.~\ref{eqPairR}-\ref{eqPairBeta} undergoes a sub-critical pitchfork bifurcation (Fig.~\ref{fig03}a) ~\cite{Strogatz2018,Crawford1991}  giving rise to a basin of attraction with a linearly stable fixed point when two particles mutually push head-on ($\alpha = \beta =\pi$, $r = 2b$, see SM~\cite{supp}).
Figures~\ref{fig02}a,b show $\alpha$-$r$ phase portraits for $\kappa b$ combinations both above and below the bifurcation (at fixed $\beta = \pi$), illustrating the formation of a basin of attraction.
A systematic bifurcation analysis of the linearized Eqs.~\ref{eqPairR}-\ref{eqPairBeta} around the fixed point (Fig.~\ref{fig03}a and SM~\cite{supp}) shows that the pair cohesion is sensitive to particles' rotation away from one another (illustrated in the eigenvectors in Fig.~\ref{fig03} b).

Remarkably, we just proved that a simple, purely geometric criterion, that does not involve any force or time scale, predicts the onset of pair cohesion, regardless of the explicit profile structure of the steric repulsion.

Previous works obtained similar equations and showed that self-propelled particles can align towards~\cite{Zhang2021b} or away from~\cite{Das2024} one another for the special case of metal-dielectric Janus particles, self-propelled by virtue of induced-charge electrophoresis, and that interact through their integrated charge distribution.
There, alignment is controlled by the sign and magnitude of the electric charge imbalance, leading to clustering~\cite{Zhang2021b} or flocking~\cite{Das2024}.
Yet, as pointed out in a recent review~\cite{Baconnier2024}, alignment in self-propelled particles is not the prerogative of electric dipoles --- rotation by an external force stems from the the generic fore-aft asymmetry from which motility of self-propelled particles originates. Below we show an important consequence of identifying the geometrical role of force-alignment where the response of a particle to the curvature of a boundary (be it a static wall, another particle, or a group of such particles) is encoded at the level of the individual particle by its curvity.

{\it Pair cohesion in vibrational robots}.
The condition in Eq.~\ref{eqAttraction} is supported experimentally by tuning $\kappa$ and $b$ in vibration-driven robots, and measuring the average pair kissing time, $\tau_k$.
Robots were built by gluing two counter-rotating vibration motors (DC Mini Vibration Motor 14000 RPM), to a PCB and connecting to a battery (LIR2477) through a switch.
The electronic circuit is then glued to a 3D printed circular chassis of a typical radius of $b_0\approx 3\;\rm{cm}$ (PLA, Prusa MK3), with a pair of soft legs (Elastic50A Resin, Formlabs), and a stiff leg (stainless steel pin, see Fig.~\ref{fig01}a and SM~\cite{supp}).
The individual robots' nominal speeds ($v_0\approx 3\;\rm{cm}/\rm{s}\approx b_0/\rm{s}$) was measured by imaging (Sony Alpha R), and tracking the robots' postitions, $x$, $y$, and orientation, $\theta$ (using standard~\cite{Allan2014,Walter2021} and homemade algorithms) as they move on a plate (Perspex), and computing the short time mean square displacement where trajectories are ballistic ($\langle \Delta r^2\rangle \approx v_0^2 t^2$~\cite{Howse2007}, see SM~\cite{supp}). Positive (blue) and negative (green) curvities ($\kappa_{\pm}\approx \pm 0.1/\rm{cm}\approx \pm 0.3/b_0$) were measured following a previously described procedure~\cite{BenZion2023,Arbel2024} (see Fig.~\ref{fig01}c and SM~\cite{supp}).
Switching between positive and negative curvity is achieved by rotating the soft-legs.
In short, an adjustable, constant, lateral force was introduced by tilting  the Perspex plate ($\vec{f} = mg \rm{sin}\varphi \hat{x}$). 
At a constant force, Eqs.~\ref{eqForce},~\ref{eqAlignment} reduce to an over-damped simple-pendulum  ($\dot \theta  = -\kappa \mu f_0 sin\theta$), which can be fitted by $\theta\left(t\right) = 2\rm{atan}\left(e^{\kappa v_0 t}\right)$ for a perpendicular initial condition ($\hat{e}\left(t=0\right)\perp \vec{F}$), which enables us to estimate the curvity of robots using inclines.

The condition for effective attraction in Eq.~\ref{eqAttraction} was then tested experimentally by dressing the robots with skirts of variable diameters: $6 - 24\;\rm{cm}$ ($1\le b/b_0 \le 4$, see Fig.~\ref{fig02}d and SM~\cite{supp}).
We placed pairs of robots facing ($\alpha=\beta=\pi$) and touching ($r=2b$) each other, turned them on, and monitored their center-to-center separation ($r$) until they no longer touch ($r>2b$), defining the kissing time ($\tau_k$).
While kissing, the robots' speed is reduced, ($|v|<v_0$), and a kissing time of a few seconds (1-10 s) is expected even in the absence of effective attraction ($\kappa b > -1$). 
But when the condition for attraction is met (Eq.~\ref{eqAttraction}), the kissing time increases by more than two orders of magnitude (Fig.~\ref{fig02}E).
This shows experimentally that the geometric condition correctly captures the attraction between a pair of skirt-wearing-robots.
We check that simulations of pairs of particles near contact display a similar diverging kissing time (see SM~\cite{supp}).

\begin{figure*}[t]
    \centering\includegraphics[width=0.8\textwidth]{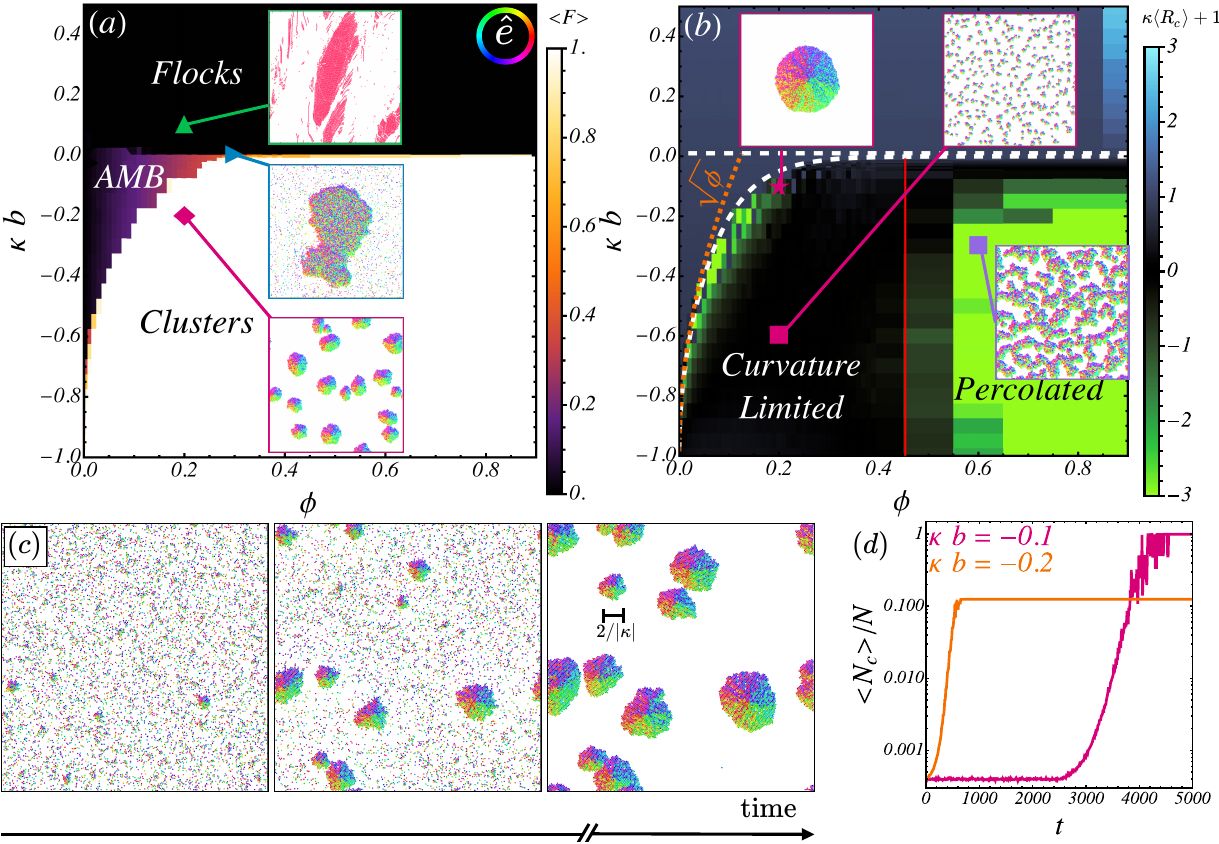}\caption{\textbf{Self-limiting clusters.}
    $(a)$ Phase-diagram ($\phi-\kappa b$) measuring the average force on each particle, $\langle F\rangle$, without noise. 
    Three types of steady states are identified (snapshots in insets; green upward triangle: flock, blue triangle: active fluid with MIPS, magenta diamond: clusters).
    In all snapshots, colors represent the orientation of $\hat{\bm{e}}$, see color wheel top-right.
    $(b)$ Intensity map of $\kappa \langle R_c\rangle + 1$ shows a valley of self-limiting aggregation (black region) where the size is curvity controlled ($\langle R_c\rangle \kappa = -1$, see Eq.~\ref{eqAttractionCluster}).
    The phases identified in $(a)$ are sketched with dashed white lines, and a $\sqrt{\phi}$ behavior at small $\phi$ is highlighted by an orange dashed line.
    A solid red line indicates the onset of percolation, at $\phi_{cp}/2$.
    Snapshots illustrate the difference between clusters with high curvature (magenta square) and low curvature (magenta star).
    $(c)$ Cluster growth dynamics leading for $\kappa b = -0.2$ and $\phi = 0.2$.
    $(d)$ Growth of the average number of particles in a cluster for $\kappa b = -0.1$ (red), and $\kappa b = -0.2$ (orange) with $\phi = 0.2$.}
    \label{fig04}
    \vspace{-0.5cm}
\end{figure*}

{\it Many-body dynamics at zero orientational noise}.
We next show that the condition for effective attraction in Eq.~\ref{eqAttraction} extends beyond zero concentration (pair-interaction), and quantitatively predicts clustering and crystallization at finite filling fractions, $\phi$.
The many-body generalization of Eqs.~\ref{eqForce},~\ref{eqAlignment} reads 
\begin{align}
    \frac{d}{dt}\vec{r}_{i} &= v_0 \hat{e}_i + \mu F_0\sum_{i\neq j} \Gamma(r_{ij})\hat{{r}}_{ij} \label{eqManyForce}\\
    \frac{d}{dt} \hat{e}_i &= \kappa \, \hat{e}_i \times \left(\vec{v}_i\times\hat{e}_i\right) + \sqrt{2 D_r}\xi_i(t) \hat{e}_i^{\perp},
    \label{eqManyAlign}
\end{align}
with $\Gamma(r_{ij}) = \min[0,k(1 - r/\sigma)]$ a harmonic repulsion with range $\sigma$ and stiffness $k$, and a unit-variance zero-mean Gaussian orientational noise $\xi_i\left(t\right)$ (such that $\langle \xi_i \rangle = 0$ and $\langle \xi_i (t) \xi_j (t')\rangle = \delta_{ij}\delta(t-t')$), with a rotational diffusion constant $D_r$ that leads to a finite persistence length, $l_P \equiv v_0/D_r$, which defines the Péclet number $\rm{Pe} = \ell_P / \sigma$. 
We set $\mu F_0 / v_0 = 1$ and $k =100$, leading to $2b = 0.99 \sigma$ and simulate Langevin dynamics of $N = 8192$ particles (see SM~\cite{supp}).

We first perform simulations at zero noise, ($1/\rm{Pe} = 0$), in 1,500 different combinations of curvities ($-1<\kappa b<0.5$), and filling fractions ($0.001\le\phi\le 0.90$) that follow Eqs.~\ref{eqManyForce},~\ref{eqManyAlign}.
Each simulation runs for a total time of $10^4 \sigma / v_0$.
The mean force on each particle, $\langle F \rangle$, reveals three distinct behaviors in the $\phi$ - $\kappa b$ phase diagram (Fig.~\ref{fig04}$(a)$).
At the top, a flocking phase where the particles have positive force alignment ($\kappa >0$) is indicated by a vanishing average force.
When curvity is weakly non-positive, we report an active fluid phase, with a force that grows continuously with concentration, and undergoes MIPS~\cite{Tailleur2008} at elevated densities, much like Active Model B (AMB)~\cite{Cates2014}.
At more negative values of curvity, an arrested, clustered crystalline phase is shown by a force that exactly compensates self-propulsion (white region).
The flocking and crystalline nature of the phases are confirmed by the mean polar $m = |\langle \hat{e} \rangle|$ and the hexatic~\cite{Nelson1979} order parameters (see SM~\cite{supp}).
At vanishing filling fraction ($\phi_{\rm{min}} = 10^{-3}$), the onset of clustering is well captured by the same geometric criterion (Eq.~\ref{eqAttraction}): $\kappa b = -1$.
With increasing concentrations, particles can become effectively attracted even at less negative $\kappa b$ values.
A cluster with an effective radius larger than the particles ($R_c>b$) becomes similarly attractive when
\begin{equation}
    \kappa + 1/R_c < 0.
    \label{eqAttractionCluster}
\end{equation}
This is illustrated in Fig.~\ref{fig04}$(b)$ where the intensity now represents the sum $\kappa \langle R_c\rangle + 1$, with $\langle R_c\rangle$ the average cluster radius (see SI): the whole clustered phase verifies $\kappa \langle R_c\rangle < -1$ (black to green intensities), with a large region that nearly saturates the inequality, so that the typical cluster size is given directly by $\langle R_c\rangle \approx -1/\kappa$. 
Using a simple kinetic theory in conjunction with  Eq.~\ref{eqAttractionCluster}, offers a clustering condition at finite concentration,

\begin{equation}
        \kappa_c b < -1 + C \sqrt{\phi_c},
    \label{eqFiniteConcentration}
\end{equation}
and quantitatively captures the boundary between the phases (orange dashed line in Fig.~\ref{fig04}$(b)$, see SM~\cite{supp}).
This criterion means that below the clustering line, the size of clusters diminishes with increasing attraction strength ($\kappa b$ more negative, see insets of Fig.~\ref{fig04}$(b)$).
Near the clustering line, the critical cluster size may be so large that a single cluster forms.
At high packing fractions, $\phi \gtrsim \phi_{cp}/2$, the system has a single percolating cluster.
A cluster growth sequence (below percolation)  shows the concluding typical cluster radius of  $\sim 1/\kappa$ (Fig.~\ref{fig04}$(c)$). 
We also show in Fig.~\ref{fig04}$(d)$ and in SM~\cite{supp} videos the time evolution of the average number of particles per cluster at two different $\kappa b$'s.
The growth is typical of non-critical nucleation-growth dynamics: clusters remain small over long times, until a critical size is reached, triggering a rapid (here, exponential) growth, similar to previously reports on accumulation of very persistent self-propelled particles~\cite{Cremer2014}).
The mean diameter of clusters is self-limited, reaching a $\kappa$-dependent plateau at steady state.

Previous work showed flocking of self-propelled particles with positive force-alignment~\cite{Deseigne2010,Das2024}, consistent with our findings: polar order is observed at the smallest positive curvity ($\kappa b = 0.005$), and at the lowest filling fraction tested ($\phi=0.001$), and is expected for a stiff potential (see SM~\cite{supp}).
By contrast, the onset of effective attraction requires a combination of finite (negative) curvity or finite concentration (Eq.~\ref{eqFiniteConcentration}).
Otherwise, the attractive fixed point in the dynamical system (Eqs.~\ref{eqPairR}-\ref{eqPairBeta}) is unstable and phase is indistinguishable from Active model B~\cite{Cates2014}: where a low density uniform isotropic liquid undergoes MIPS at a higher density, even at zero noise~\cite{ Solon2015,Nie2020,Casiulis2021}.
It was previously speculated that the negative alignment will show an enhanced Motility Induced Phase Separation (MIPS) as seen in Active Model B~\cite{Baconnier2024}. However, we find the existence of a threshold beyond which negative alignment leads to a new phase behavior: particles self-assemble into solid-like clusters with self-limiting size.
Once multiple particles form a sufficiently large cluster, other particles become effectively attracted to the cluster. Similar to the pair attraction criterion (Eq. 1), force aligning active particles (Eqs.~\ref{eqForce},~\ref{eqAlignment}) become attracted to a stationary repulsive potential provided that its radius is sufficiently large~\cite{Arbel2024}. Treating the cluster as a stationary obstacle of radius $R_c$, the single particle criterion for attraction naturally extends to the many-body system (Eq.~\ref{eqAttractionCluster}). The self-limiting clustering observed with negative alignment is categorically different than the fluid-fluid co-existence found in MIPS.
The existence of a threshold for effective attraction (Eq.~\ref{eqAttraction}) may explain why the novel self-limiting aggregation was not observed in prior work.

{\it Phase behavior at finite noise}.
\begin{figure*}[t]
    \centering
    \includegraphics[height=0.28\textwidth]{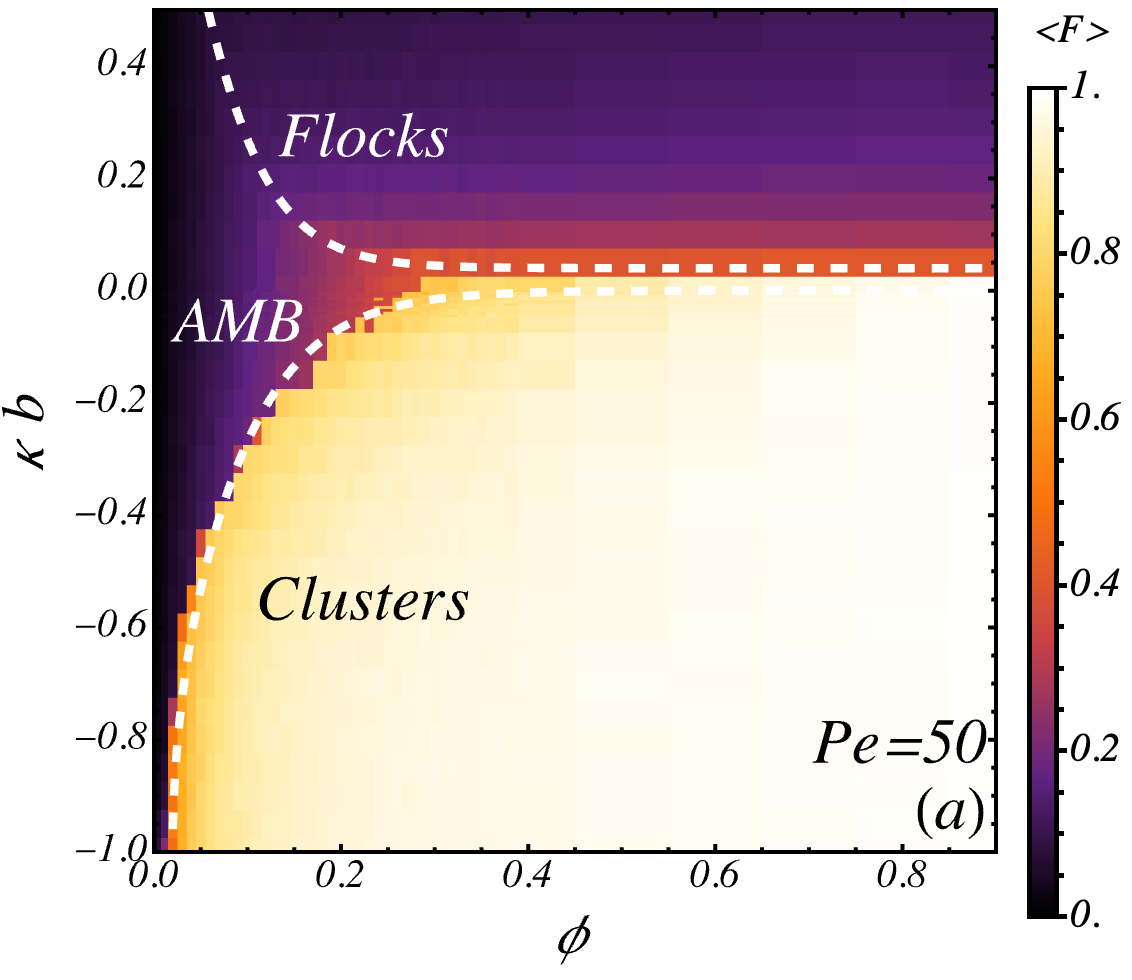}
    \includegraphics[height=0.28\textwidth]{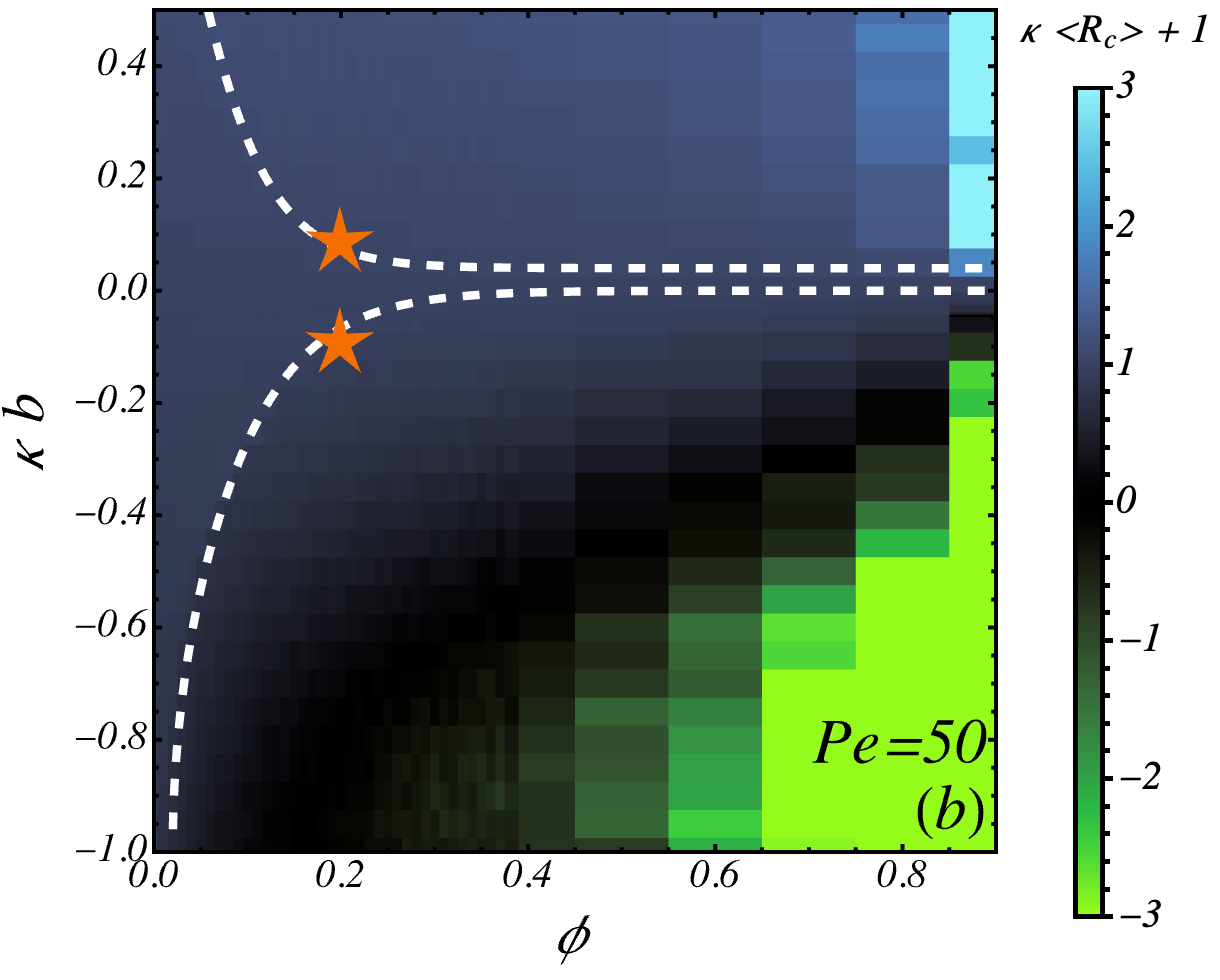}
    \includegraphics[height=0.28\textwidth]{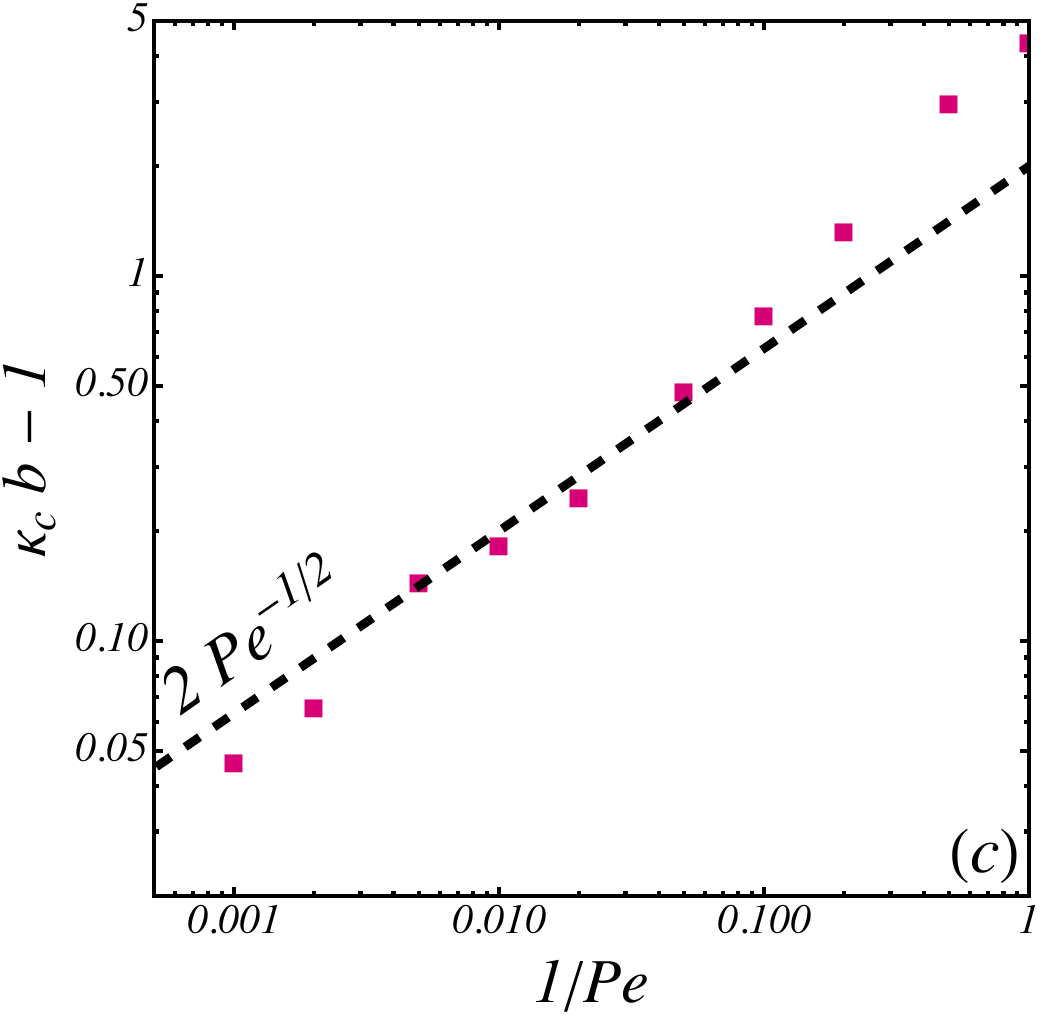}\\
    \includegraphics[width=0.9\textwidth]{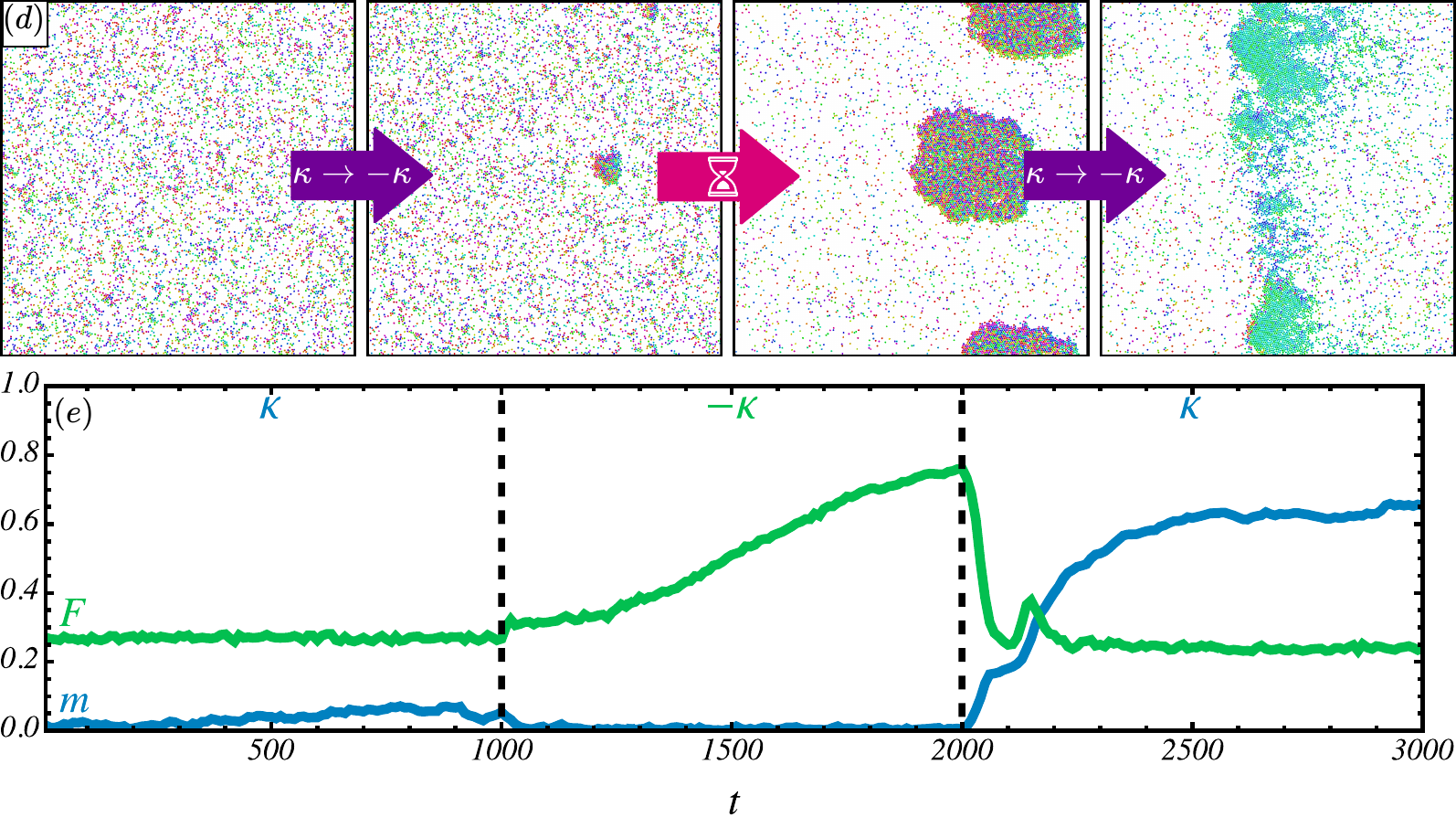}
    \caption{\textbf{Self-limiting aggregation and decentralized control at non-zero noise.}
    $(a)$ Intensity map of the mean force felt by particles in the $\phi, \kappa b$ plane at $Pe = 50$ and $(b)$ corresponding $\kappa \langle R_c \rangle$ map.
    Dashed white lines indicate phase boundaries.
    $(c)$ Asymptotic zero-density $\kappa_c b -1$ against $1/Pe$, in log scales.
    A dashed line shows $ 2 Pe^{-1/2}$ (Eq.~\ref{eqAttractionNoise}).
    $(d)$ Decentralized control strategy: switching $\kappa$ to $-\kappa$ triggers fast clustering, then long-lived metastable flocking as $\kappa \to -\kappa$.
    $(e)$ Example curves for magnetization, $m$ (blue), and average force, $F$ (green), with switches at dashed black lines.
    Parameters for $(d)$ and $(e)$ are shown as orange stars in $(b)$.
    }
    \label{fig05}
    \vspace{-0.5cm}
\end{figure*}
So far we discussed deterministic dynamics with zero orientational noise ($D_r = 0$ in Eq.~\ref{eqManyAlign}).
With no fluctuations, the system settles when mechanical force balance is achieved  --- the clusters are arrested.
In this final section, we show that the observed phases are stable even when particles have orientational diffusion and that the picture portrayed by the geometric construction is valid at finite noise.

Figure~\ref{fig05} shows that the noiseless phase diagram is qualitatively preserved, even in the presence of fluctuations --- the same 3 distinct phases are observed even though the dynamics are no longer arrested (see Videos in SM~\cite{supp}).
This is also captured quantitatively: while the AMB-clustering phase boundary remains sharp, the average force on each particle is too low to ensure mechanical equilibrium ($\mu \langle F \rangle /v_0 < 1$).
As before, the same phase boundary is observed when measuring the polar and hexatic order parameters (see SM~\cite{supp}).
A similar picture is also painted for the curvity limited cluster size (Fig.~\ref{fig05}$(b)$): when the curvity is not sufficiently negative to satisfy the condition for pair cohesion (Eq.~\ref{eqAttraction}) but the concentration is sufficiently high, the typical cluster size is again $R_c \approx -1/\kappa$ (Eq.~\ref{eqAttractionCluster}).
Fluctuations allow individual particles to adhere to or detach from a given cluster, yet the typical cluster size is maintained at steady-state, in a dynamical equilibrium (see Video in SM~\cite{supp}).
The onset of attraction at the infinite dilution limit is shifted in the presence of noise, but the shift is also quantitatively captured by extending the geometric criterion to finite persistence lengths (see SM~\cite{supp}),
\begin{align}
    \kappa + \frac{1}{b} + \frac{2}{\sqrt{b\ell_p}} < 0.
    \label{eqAttractionNoise}
\end{align}
Figure~\ref{fig05}$(c)$ shows that the $1/\sqrt{l_p}$ ($1/\sqrt{Pe}$) scaling above quantitatively captures the onset of attraction at infinite dilution for over 2 orders of magnitude of orientational noise ($0<1/Pe <0.1$) then starts growing faster when the particles' persistence approaches their own size ($1/Pe>0.1$).
This extended criterion can be understood by noticing that the stochastic term encodes the typical curvature in the trajectory of a free particle (see SM~\cite{supp}).

The above results offer a microscopic handle for programmable self-assembly of self-limiting clustering, with important applications in the design of mechanical~\cite{Bertoldi2017} and photonic~\cite{Joannopoulos2008} meta-materials, and in living matter~\cite{Hagan2021}.
We further show its application as a novel control architecture for a multi-agent robotic system. 
Real-world flocking applications require robustness in a sparse and noisy environment, where a stable flocking phase is not expected, even at significant curvities ($0<\kappa b \approx 1$, Fig.~\ref{fig05}a).

Since the curvity and a particle's radius are dependent~\cite{Arbel2024,Baconnier2024}, inducing flocking by arbitrarily increasing $\kappa b$ is not always physically feasible.
We propose toggling the curvity's sign as an alternative route (while roughly keeping its magnitude).
This was achieved with the robots above by switching the orientation of their soft-legs (see Fig.~\ref{fig01}$(a)$). Adding a servo, pneumatic actuator or an electro permanent magnet to the current design can change the orientation of the soft legs (by rotating or bending) to facilitate an embedded curvity control.
A sequence of two curvity flips can lead to flocking even at globally low concentrations and finite persistence length: starting from a disordered fluid (low $\phi$ and $\kappa >0$) switching to $-\kappa$, induces an effective attraction, and most of the particles spontaneously cluster.
With the elevated local density inside the clusters, switching back the curvity sign ($\kappa>0$) induces alignment, and the swarm spontaneously flocks (Fig.~\ref{fig05}d and the corresponding video in SM~\cite{supp}).
Following the curvity sign-flip the exponential cluster growth expedites the flocking, as also shown by the growth of average force and average magnetization modulus in Fig.~\ref{fig05}$(e)$.

{\it Conclusions}. 
In this work we presented a geometric criterion for the onset of attraction and effective cohesion between pairs of robots, which depends on the coupling between two intrinsic properties of self-propelled particles: 1. their morphological curvature, $1/b$ (inverse radius), and 2. their curvity, $\kappa$: the signed, charge-like property of self-propelled particles that characterized the curving of their trajectories when subjected to an external force.
The criterion (Eq.~\ref{eqAttraction}) shares mathematical structure with the Young-Laplace equation~\cite{DeGennes2002}, where the stability of a three-dimensional fluid interface is conditioned by the sum of two local curvatures, suggesting a link between interfacial phenomena, boundaries, and active matter.
We showed that these can be designed with real robots, and found experimentally and numerically that the geometric criterion predicts the onset of effective pair attraction.
Interestingly, the phenomenology closely resembles that observed in Janus colloids subjected to electric fields~\cite{Zhang2021b,Das2024}, suggesting that $\kappa$ is a mechanical analog of the charge imbalance of such particles.
The geometric criterion obtained in the infinite dilution limit extends to finite concentrations and finite noise, and explains the typical cluster size observed, offering a powerful microscopic rule for tuning a macroscopic length scale, with important applications in material science and in living matter.
We proposed how this construction can be used as a new control paradigm in multi-agent robotic systems, its broader applicability in dry active matter, and implication for biological and robotic swarms.

{\it Acknowledgements}. 
M.C. would like to thank Satyam Anand for insightful discussions on this work.
M.C. and S.M. acknowledge the Simons Center for Computational Physical Chemistry for financial support.
This work was supported in part through the NYU IT High Performance Computing resources, services, and staff expertise.
M.Y.B.Z. acknowledges the support of the Dutch Brain Interface Initiative (DBI2), project
number 024.005.022 of the research programme Gravitation financed by the Dutch Research Council (NWO).
%


    

\bibliography{clusterFlock,supp}

\end{document}


\maketitle

\SItext

\renewcommand{\figurename}{FIG.}
\renewcommand{\thefigure}{S\arabic{figure}}
\renewcommand{\thetable}{S\arabic{figure}}

\newtagform{S}{(S})
\usetagform{S}

\begin{center}
{\bf Supporting Information}    
\end{center}

\section{\label{secExpSI} Experimental Methods}
%
\subsection{Building Robots}
Home-made robots (Fig.1a in main text) were built following a similar procedure as described in previous work~\cite{Arbel2024}. 3D printing files are found in the supplementary information.\\

\textbf{Parts include:}

\begin{figure}[H]
    \centering
    \includegraphics[width=0.2\linewidth]{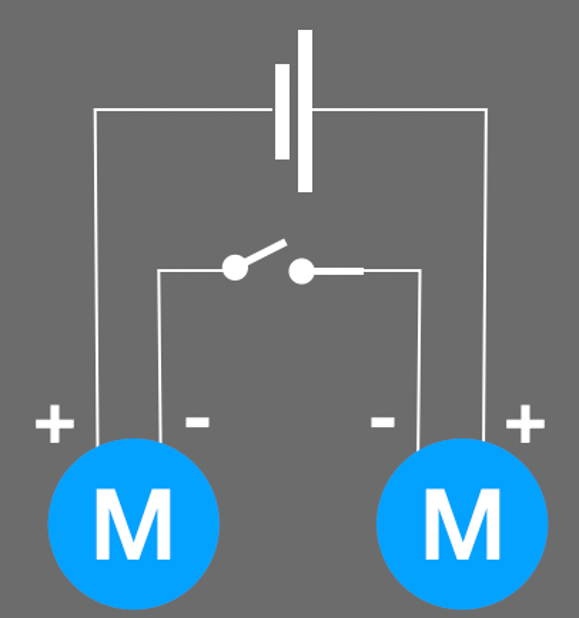}
    \caption{\textbf{Robot's electric circuit}}
    \label{fig:circuit}
\end{figure}

\begin{itemize}
\item \textbf{Chassis:} Made with PLA filament and in the printer Bambu Lab x1, diameter of 60 mm
\item \textbf{Soft legs:} Made with SLA printer of Formlabs with Elastic 50A Resin. The size is 10X5mm
\item \textbf{Stiff leg:} Metallic cylinder
\item \textbf{Motors:} DC vibration motors of BestTong 6mmX14mm with 14000 RPM.
\item \textbf{Motors' circuit:} The circuit has been soldered manually. We used PCB and a buttery house which fit to a LIR2477 battery, two motors and a switch.The soldering was made according to the drawing \ref{fig:circuit}. Then motors were glued to the PCB with UV glue.

\item \textbf{Robots' skirts:} Made with PLA filament and in the printer Bambu Lab x1, with the diameters of 98 mm, 114 mm, 131 mm, 196 mm and 230 mm.

\end{itemize}

\textbf{Assembly steps:}
\begin{enumerate}
\item Glue PCB to the chassis grove.
\item Glue soft legs
\item Glue the stiff leg temporarily.
\item Tune the height of stiff leg according to the robot's straight track.
\item Permanently glue the stiff leg.

\end{enumerate}

\subsection{Imaging}

Video acquisition was performed using a Sony Alpha 7S camera with a Sony FE 12-24 f/4 G lens at a frame rate of 25 frames per second, and saved in an mp4 format. Robots' position and orientation were extracted using the interactiveLocate.ipynb python notebook available on github.

\subsection{Position and Orientation Tracking}
The tracking of the position with the orientation of the robot done using Trex~\cite{Walter2021}. The method was to save the position of the chassis's center and the position of the battery's center. From these positions and according to the known structure of both kinds of robots we extract the orientation vector of the robot on each frame.

\subsection{Measuring Robots' Nominal Speed}
Robots' nominal speed ($v_0$) was extracted from the mean square displacement (MSD) of their trajectories. This was performed by locating the robots' positions, linking their positions into trajectories and then computing the MSD. Fitting the short time MSD where the motion is ballistic (MSD $\propto v_0^2\tau^2$), gives the nominal speed. This was done on logarithmic scale where:
\begin{equation}
    \log MSD\approx   \log \tau^2 + \log v_0^2 
\end{equation}
where $v_0$ is the nominal speed  and $\tau$ is the lag time.
\\

\subsection{Measuring Curvity}

Robots' curvities ($\kappa$) were measured by tracking their orientation while moving on an inclined plane of variable slope. The plane used for the measurement was made of acrylic (identical to the material of the arena used in the other experiment). The measurements done with one fronter robot and one aligner robot (see supporting videos). The angles used were $0.8^\circ$, $2.3^\circ$, $4.4^\circ$, $5.4^\circ$, $7.3^\circ$, $9.4^\circ$, $11.5^\circ$ as measured using a digital tilt measurement tool (INSIZE  2173-360), accurate to within $0.5^\circ$. The number of repetition for each angle's slope range between 7-15.

Videos were analyzed by extracting the robot's chassis center and battery center, from which its orietnation was computed as a function of time $\theta \left(t\right)$ relative to the x-axis, which was then fitted for each measurement according to the relation:
\begin{equation}
   \tan\frac{\theta}{2}=e^{At} 
\end{equation}
Where $A$ present the expression $A= \kappa mg \mu\sin\varphi$, and $\varphi$ is the angle of the slope.
Later, the  $<A>$ for each slope was computed by averaging the measurements of the repetitions and then extracting with another fit $\kappa$ according to representation relation of $A$.

\subsection{Measuring Kissing Time}
Kissing times were measured by initializing pairs of identical robots osculating and facing each other, turning them on and monitoring the time over which they are in contact. Contact was defined as long as their center to center separation was less than their summed radii plus a 5 pixel margin. Experiments with each pair type was repeated 10-17 times.
Kissing time experiments were limited to 10 minutes, ensuring a consistent power from the battery.

\subsection{Mechanical Origin of Force Alignment in granular hoppers}

We present a step-by-step derivation of the mechanical parameters that control the robot's curvity (following the derivation found in~\cite{Arbel2024}). The rapid vibrational motion of the robots is captured by three phases (see inset of Fig. 1a in main text): (i) an aerial phase where the robot is completely aloft; (ii) a pivot phase, where the robot spends time on a soft leg; (iii) a resting phase, where the robot recovers before the next hope. The combined duration of the three phases is on average $T$. The motion is assumed to be quasi 2D, static friction at contact, with complete loss of momentum upon landing, and the robot experience an external lateral body force acting in the plane of motion, $\vec{f}$. The inertial dynamics of the rapid motion are described by the robot's instantaneous linear and rotational positions ($\vec{r},\theta$), speeds ($\dot{\vec{r}},\dot\theta$), and accelerations ($\ddot{\vec{r}},\ddot\theta$), where $\theta$ defines the robot's heading ($\hat e \equiv \left(\cos \theta,\sin\theta\right)$) and is measured counter-clockwise relative to the $x$ axis. 

\begin{enumerate} 

	\item During the \textbf{Aerial phase (i)}  the robot springs forward with an instantaneous horizontal speed of $v_h$, and stays aloft for the aerial time $\tau_A$. The mean displacement the robot experiences during this phase is $\langle \Delta R\rangle = v_h \tau_A \hat {e} + \tau_A^2/2m \vec{f}$. While in the air, the robot does not experience a torque, and the mean rotation is zero $\langle \Delta \theta \rangle = 0$. Averaged over the mean duration of the three phases ($T$) gives 

$$ \langle \dot {\vec{r}} \rangle \equiv \frac{\langle \vec{r} \rangle}{T} = \vec{v} = v_0\hat e +\mu \vec{f},$$

 where the effective parameters are now defined according to the microscopic properties: the mobility: $\mu \equiv \tau_A^2/2mT$, and the nominal speed: $v_0\equiv \tau_A/T v_h$. This is Eq. 2 in the main text.

	\item During the \textbf{Pivot phase (ii)} the robot has zero linear acceleration (to leading order), therefore $\langle \Delta \vec{r}\rangle \approx 0$. Since it contacts the floor at one point, the robot experiences a torque ($\vec \delta \times \vec f$), where the lever arm is given by the displacement of the Center of Mass (CoM) from axis of rotation along the heading ($\vec\delta \equiv \delta \hat e$, see inset of Fig.1a in the main text). During the pivot time ($\tau_P$), the robot experiences an average rotation of $\langle \Delta \theta \rangle \hat z = \delta \tau_P^2/2I \hat e \times \vec f$. Averaging over the hopping sequence, the mean rotational speed is
	
	$$ \langle \dot \theta \rangle \hat z \equiv \frac{\langle \Delta \theta \rangle}{T}\hat z = \vec{\omega} = \frac{1}{2}\frac{\delta \tau_P^2}{I} \hat e\times \vec f.$$
When expressed using the heading vector, this becomes:

$$\dot{\hat e} = \kappa \hat e \times \left(\vec v\times \hat e\right)$$,

where the cruvity is defined from the microscopic parameters independent of the horizontal speed: $\kappa \equiv m\delta/I \left(\tau_P/\tau_A\right)^2$.

	\item The \textbf{resting phase (iii)}, is the recovery phase where the robot contacts the surface in more than one point and both linear and rotational accelerations are zero.
	
	\end{enumerate}

\subsubsection{Evaluating Robot's Curvity From Mechanical Design}

In order to evaluate the curvity $\kappa$ of the developed robots the design's mass distribution was assumed to consist of a thin circular chassis (where mass is concentrated at its perimeter) and a dense core (where the mass is equally distributed in a disc). The chassis is predominately made of the 3D printed material, and the core is made by large of the battery and some associated electronics. We define the overall mass $M$, the mass of the core $m_B$ (with a radius $R_B$, and the mass of the chassis $m_C$ with a radius $R_C$. The moment of inertia of the core around an axis at its center is estimated to be that of a disc $I^0_B = 1/2m_C R_B^2$ and the moment of inertia of the circular chassis around its center is estimated to be that of a thin ring $I^0_C = m_C R_C^2$. Since the pivot point is not at the center of the core nor the center of the chassis, we use the parallel axis theorem to evaluate the moment of inertia around the displaced axis $I_C = I^0_C + m_C r_C^2$ and $I_B = I^0_B+m_B r_B^2$ where $r_C$ and $r_B$ are the displacements of the pivot axis from the center of the chassis and the core respectively.  Finally, the over all moment of inertia of a robot for rotating around the pivot point is given using the principle of superposition: $I = I_C + I_B$.

Using the above relations, along with the designed geometries and measured masses (see Tables I and II, as well as design files) we find a moment of inertia of $I_H = 448$ g cm$^2$.

The next set of parameters required to evaluate the curvity is the aerial and pivot times ($\tau_A$ and $\tau_P$ respectively). An accurate evaluation requires a high speed measurement of the subtle contact of the rapidly vibrating robots, but can be estimated to be roughly equal ($\tau_P/\tau_A\approx 1$) (see Arbel 2024 et al, ref 58 in the main text) simplifying the evaluated curvity to $\kappa \approx \frac{M}{I}\delta$.

Using the center of mass displacement $\delta\approx 2.5$cm and the overall robot's mass $M\approx 32 $g, along with the above calculated moment of inertia, we find that the expected magnitude of the curvity is $\kappa \approx 0.18$ cm$^{-1}$, 80\% higher than value measured using the inclined plane experiments.


\section{\label{secSimulationSI} Numerical Simulations}

\subsection{Details of Simulation condition}

All the results presented in the main text are obtained via molecular dynamics (MD) simulations with the simplest possible order-1 integrator.
Namely, we write the equation of motion of any DoF a particle symbolically as
\begin{align}
    dx = x(t+dt) - x(t) = v_{det} dt + v_{stoch}dt^{1/2},
\end{align}
where $dt$ is a fixed time step, $v_{det}$ is the deterministic part of the update, that comes from self-propulsion and interactions with other particles, and $v_{stoch}$ is the stochastic part of the update that appears when we introduce noise.
In the case with noise, the stochastic part of the velocity simply reads $v_{stoch} = \sqrt{2 D_r} \eta_x$, with $\eta_x$ drawn from a unit-variance centered normal distribution, and it is zero otherwise.
The computation of the interaction part of $v_{det}$ is accelerated by introducing a partition of space into square cells twice as wide as the longest-range interaction in the system, and labelling at all times each particle with its cell number.

To set the filling fraction, in practice, we set $\sigma$ the full repulsive diameters of particles to $1$ and adjust the sidelength $L$ to achieve $\phi = N \pi \sigma^2 / (4 L^2)$, where $N = 8192$ throughout the paper.
The repulsion is chosen to be harmonic, $F(r) = \min[0, k( \sigma - r)]$ with a $k$ such that $k \sigma / v_0 = 100$.
Time is counted in units of $\sigma / v_0$, and we set the time step to, at most, $dt = 10^{-4}$, or to the largest power of ten that ensures that no update $dx$ can be larger than $0.01$ in simulation units.
The initial positions of particles are each drawn uniformly in a periodic square simulation box with linear size $L$, and the initial polarities are drawn uniformly on the circle.
Each simulation is run until the simulation time reaches $10^4 \sigma / v_0 \geq 10^8 dt$.\

\subsection{Pairwise measurements}

Kissing times are obtained numerically by simulating noiseless dynamics of pairs of particles, initially located at $\bm{r}_1 = (0,0)$ and $\bm{r}_2 = (\sigma,0)$, and with headings $\theta_1$, $\theta_2$ randomly drawn from uniform distributions in $[-\delta, \delta]$ and $\pi + [-\delta, \delta]$, respectively.
We choose $\delta = 0.1 \text{rad}$ to mimic experimental uncertainties.
The kissing time is measured as the first time at which the two particles stop interacting forever, which is obtained from the condition $r_{12} > \sigma$ as well as their headings not leading to a future collision, which may be expressed from their positions and velocities like in a conventional event-driven simulation of hard discs~\cite{Frenkel2001}.

\begin{figure}[H]
    \centering
\includegraphics[width=0.5\linewidth]{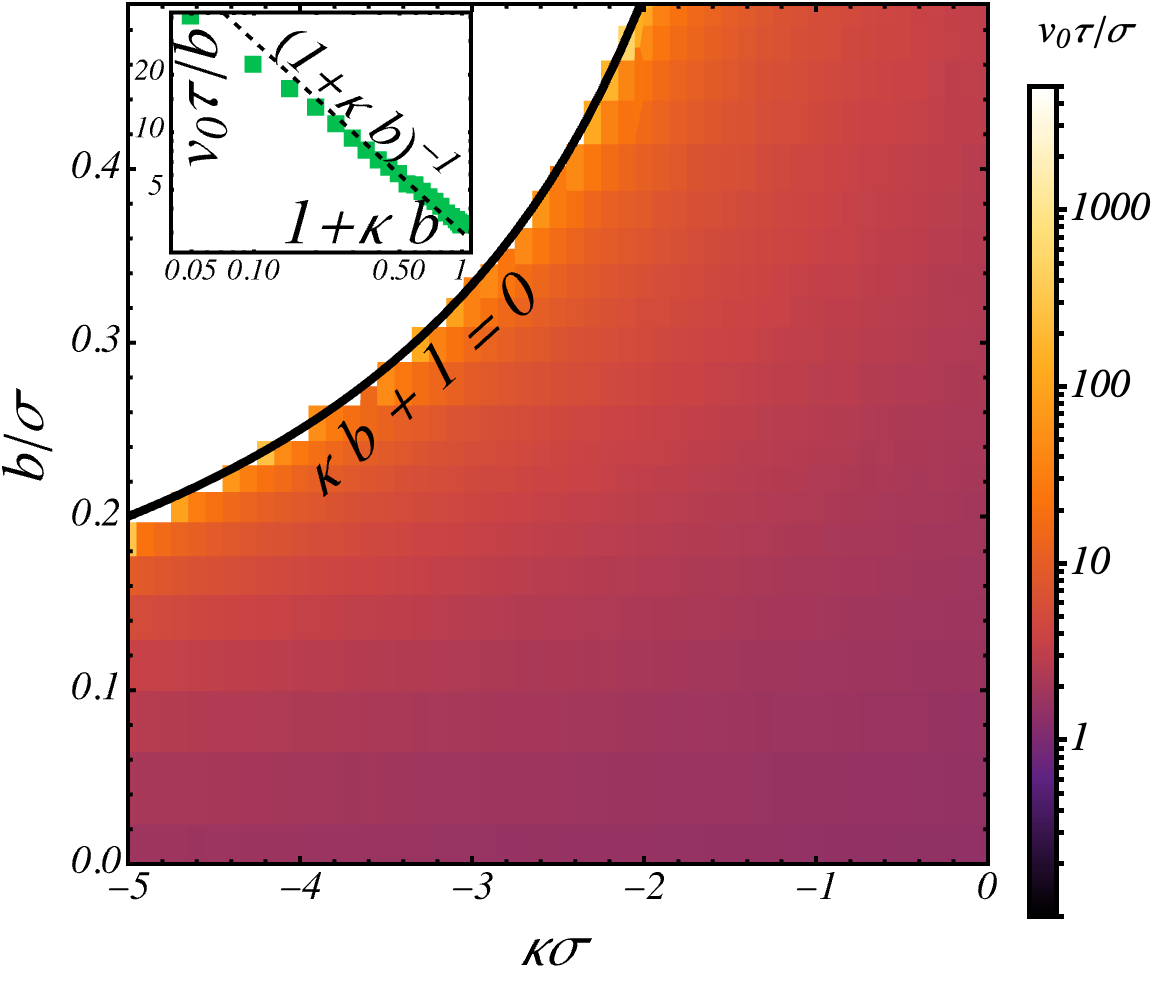}
    \caption{\textbf{Kissing times of pairs of particles.}
    Main panel: intensity map of the kissing time as a function of the dimensionless curvity $\kappa \sigma$ and equilibrium distance $b/\sigma$ between particles.
    A solid black line indicates the geometric criterion for attraction.
    Inset: Lineplot of the dimensionless kissing time $v_0 \tau / b$ against $1 + \kappa b$ for one specific choice of $b$, in log-log scales.
    A dashed line indicates a $(1+\kappa b)^{-1}$ trend.
    }
    \label{fig:kissing}
\end{figure}

The results are shown in Fig.~\ref{fig:kissing}.
In the main panel, we vary the curvity and equilibrium distance between particles (through the stiffness of the repulsion), and report the kissing times as a log-scale intensity map.
We show that the kissing time diverges as the criterion is approached.
Furthermore, in the inset, for a specific choice of $b = 0.99 \sigma$, we report a line plot of the dimensionless kissing times $v_0 \tau / b$ against the distance to the geometric onset of attraction, $1 + \kappa b$.
The result suggests a $(1+\kappa b)^{-1}$ divergence of the kissing time: this scaling is consistent with the minimal kinetic theory introduced in Sec.~\ref{sec:kinetic} to justify the finite-density evolution of the geometric criterion.

\subsection{Many-body measurements}

In simulations, we measure a number of averaged quantities.
First, in the main text, we show measures of the average force felt by a particle, given by the average over all particles of 
\begin{align}
    F = \sum_{j \neq i} \min[ 0, k (\sigma - r_{ij})].
\end{align}
In this section, we also show maps of the ``magnetization'' modulus $m$ of the system, defined as the average polarity
\begin{align}
    m \equiv \left\| \frac{1}{N} \sum\limits_{i=1}^{N} \hat{\bm{e}}_i \right\|,
\end{align}
as well as maps of the average modulus of the hexatic order parameter~\cite{Nelson1979},
\begin{align}
    \langle |\psi_6|\rangle = \frac{1}{N} \sum\limits_{m = 1}^{N} \left| \frac{1}{|\partial m|}\sum\limits_{n \in \partial m} e^{6 i \varphi_{mn}} \right|,
\end{align}
where the inner sum is taken over the $|\partial m|$ Voronoi neighbors of particle $m$, and $\varphi_{mn}$ indicates the angle formed between the vector $\bm{r}_{mn}$ and the horizontal.
The choice of taking the average of the modulus, not the modulus of the average, makes it such that this quantity is large if particles are in locally $6$-fold symmetric environments, even if there might be several orientations of the $6$-fold order in the whole system (in other words, this order parameter is high if the system consists of a large number of crystallites with different orientations).

\begin{figure}[t]

\includegraphics[width=0.21\linewidth]{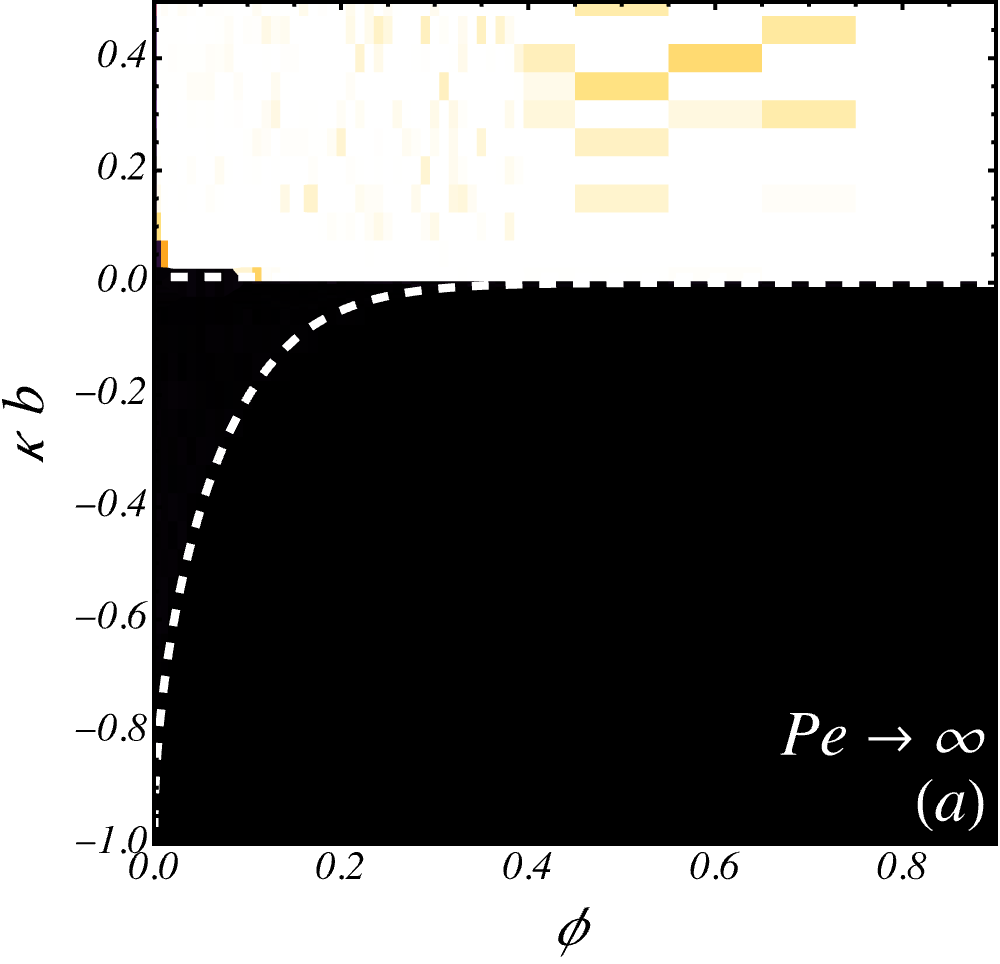}
\includegraphics[width=0.21\linewidth]{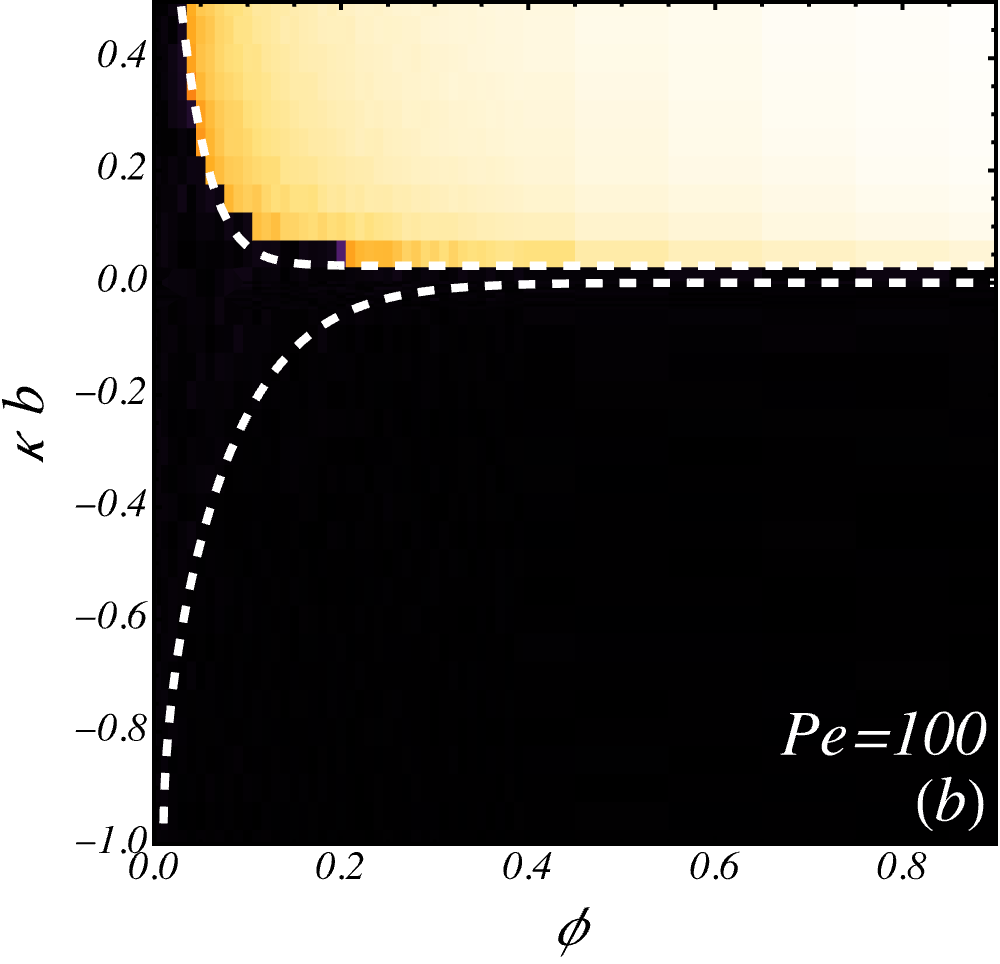} 
\includegraphics[width=0.21\linewidth]{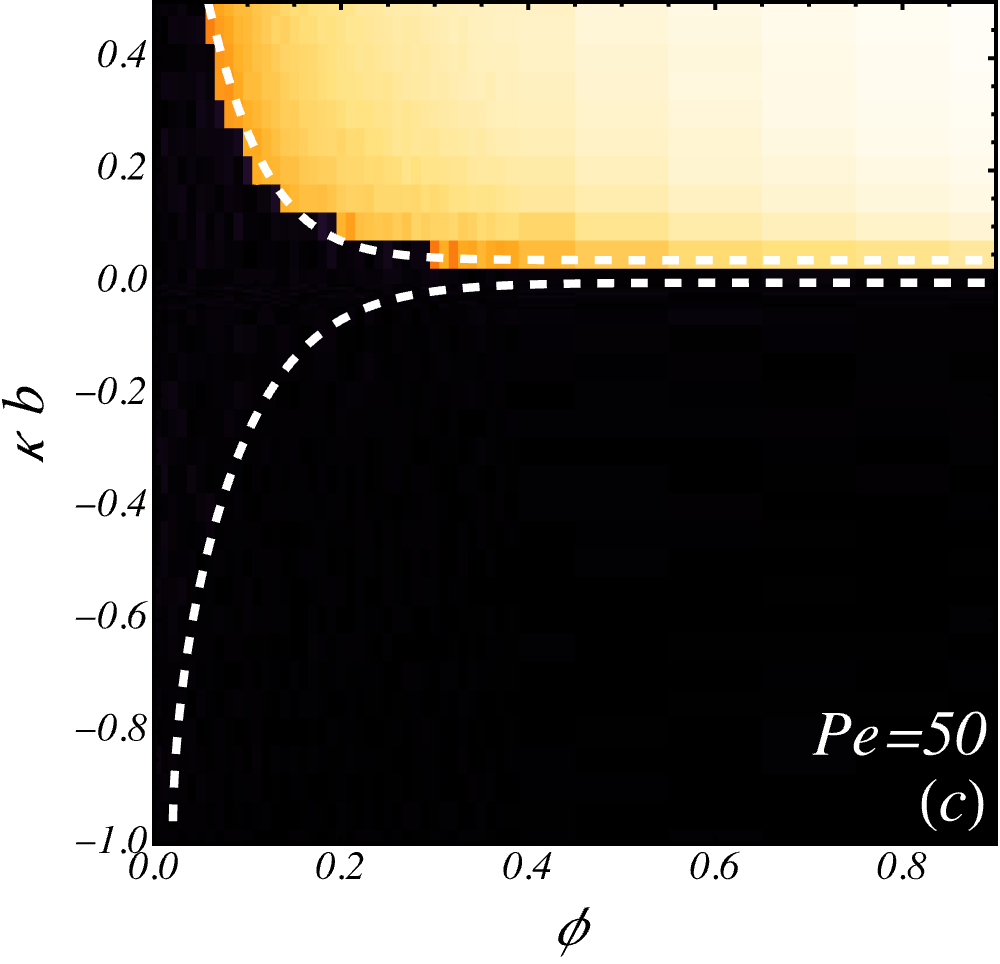} 
\includegraphics[width=0.21\linewidth]{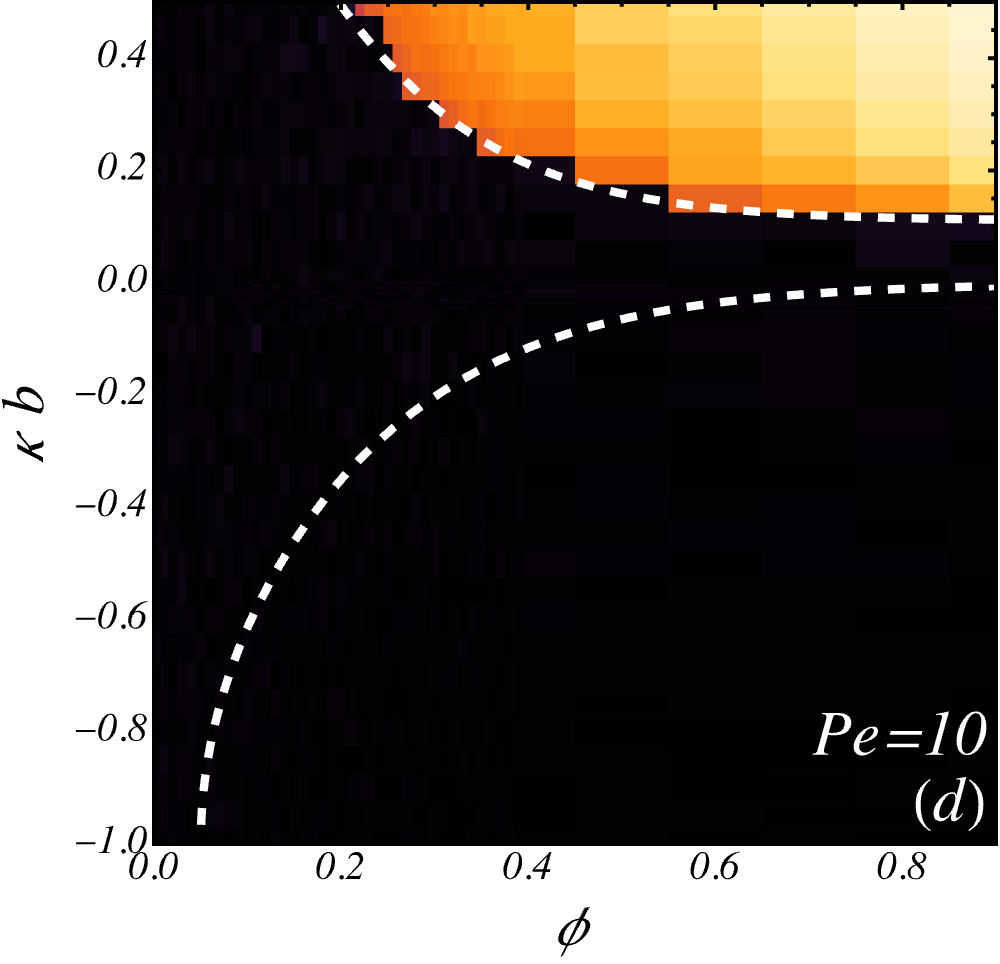} 
\includegraphics[height=0.21\linewidth]{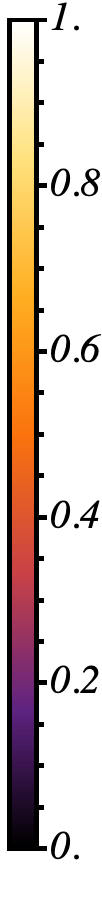} \\
\includegraphics[width=0.21\linewidth]{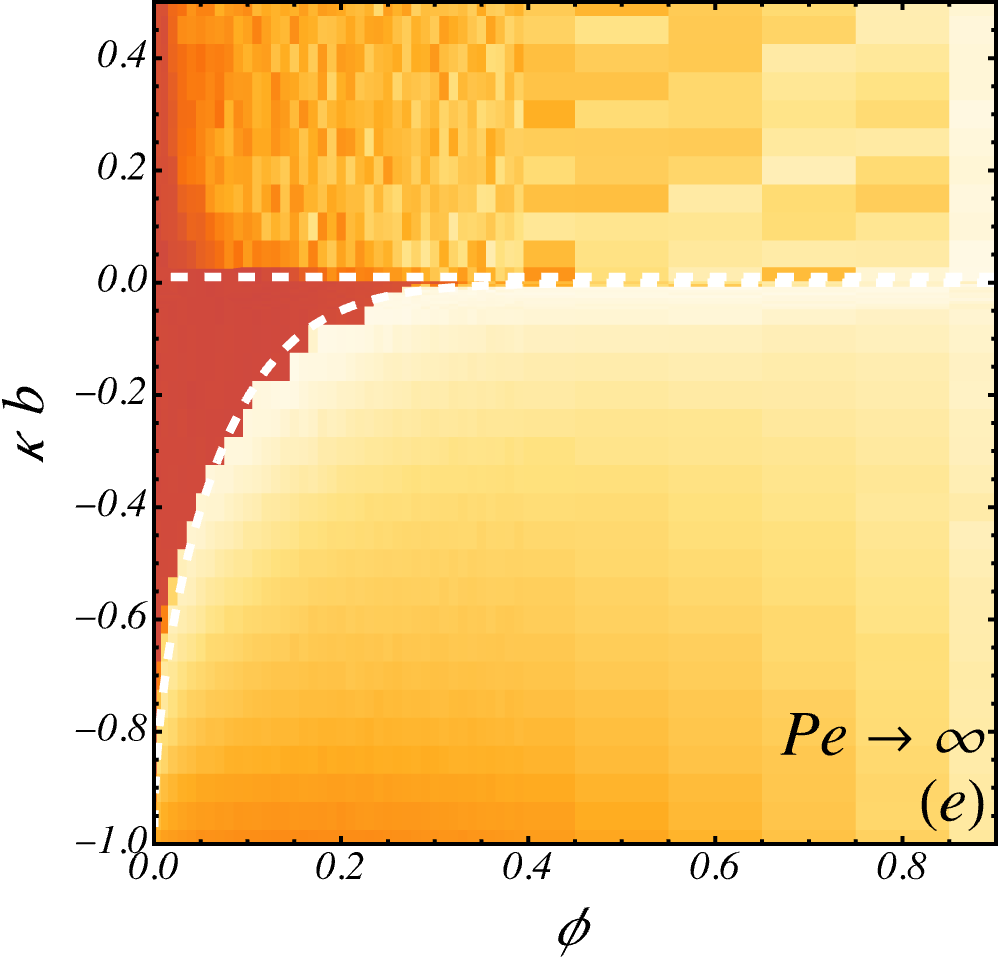}
\includegraphics[width=0.21\linewidth]{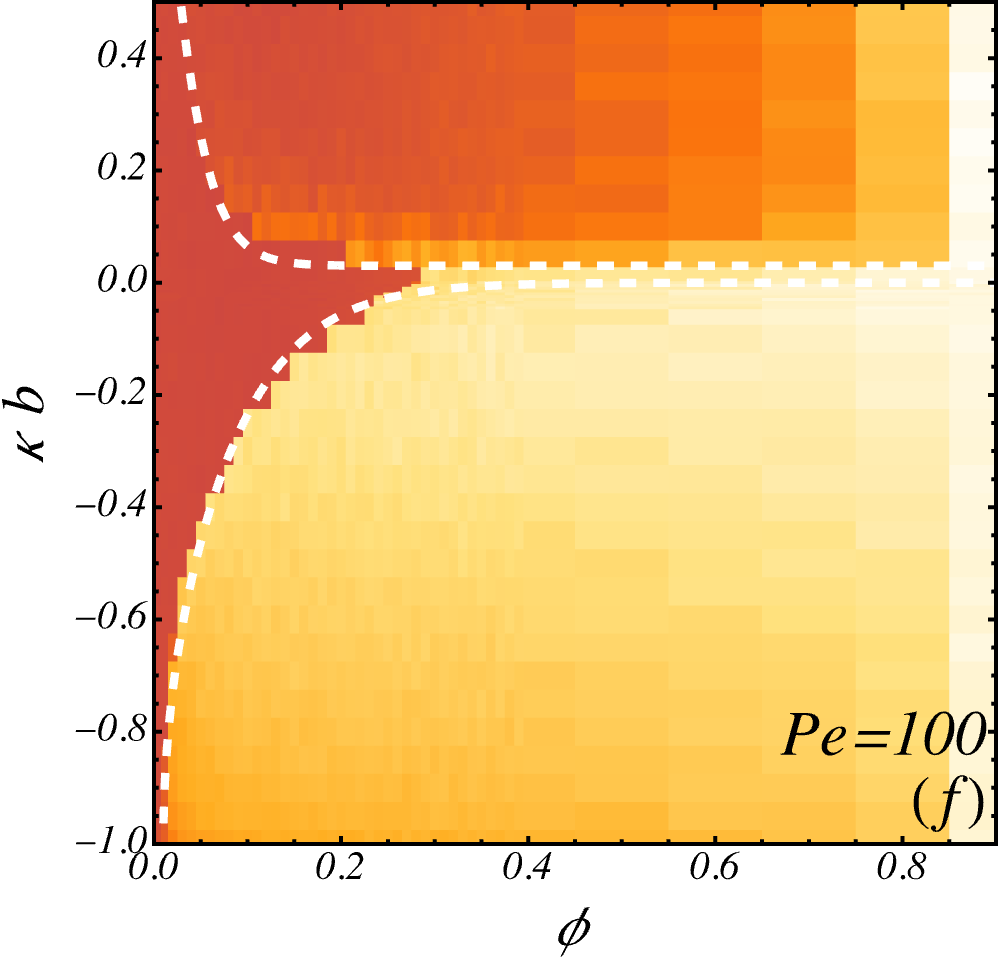} 
\includegraphics[width=0.21\linewidth]{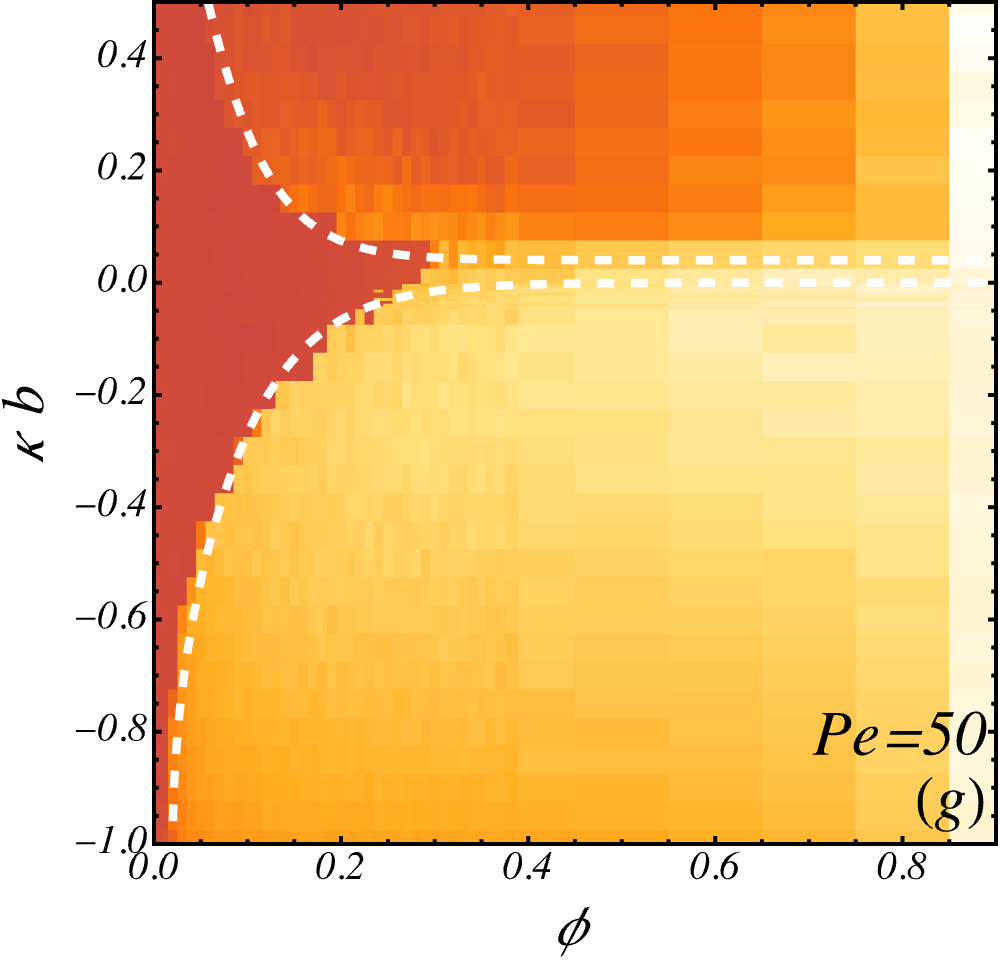}
\includegraphics[width=0.21\linewidth]{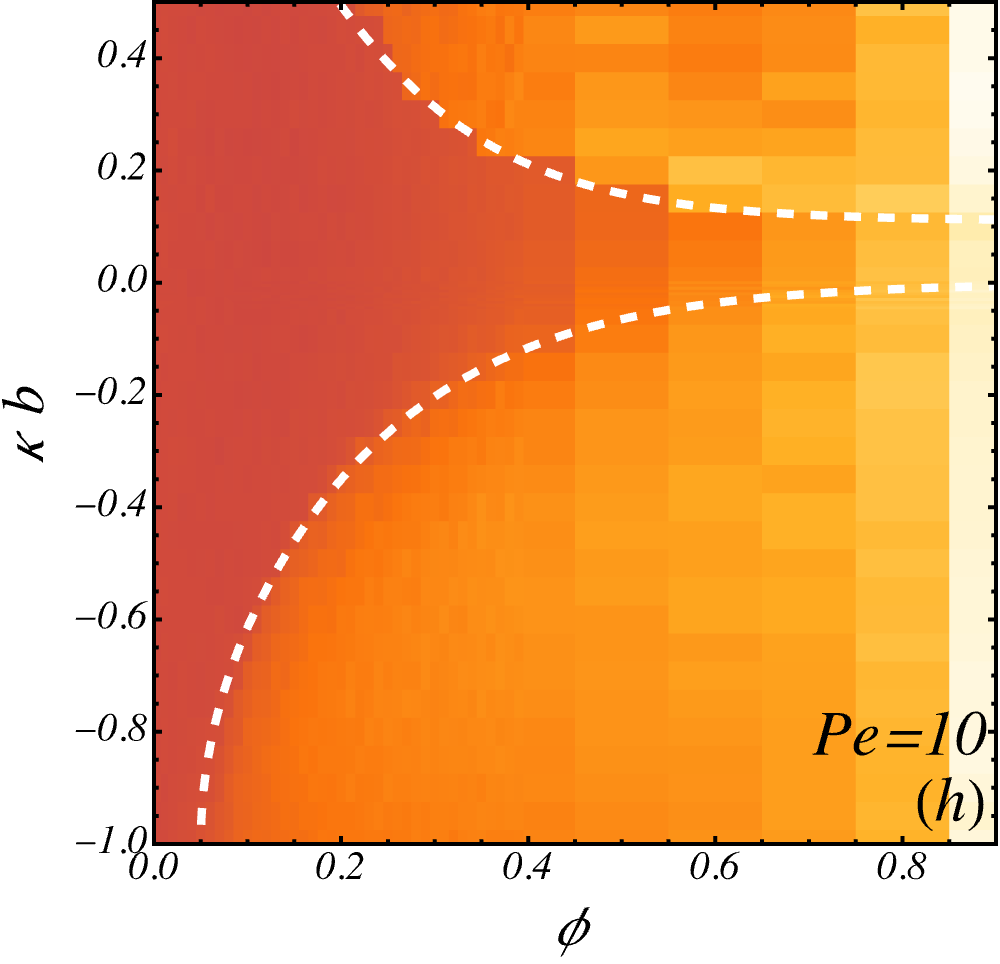}
\includegraphics[height=0.21\linewidth]{Figures/supporting/ExtraDiagrams/colorbar.png} 
    \caption{\footnotesize{Additional maps of order parameters.
    The top row shows $m$ at $(a)$ $1/Pe = 0$, $(b)$ $0.01$, and $(c)$ $0.1$, while the bottom row shows  $\langle | \psi_6 |\rangle$ at $(c)$ $1/Pe = 0$, $(d)$ $0.01$, and $(e)$ $0.1$.
    The dashed white lines show estimated boundaries for the different phases described in the main text.
    }}
    \label{figComparingOrderParams}

\end{figure}

Finally, in the main text, we introduce the cluster size $R_c$.
In practice, it is measured through the number of particles $N_c$ belonging to each cluster, defined as collections of particles connected by at least one non-zero repulsive force.
From $N_c$, the radius is estimated by noticing that the particles are close-packed inside clusters, so that the packing fraction inside is $\phi_{cp} = \pi / \sqrt{12} \approx 0.906$.
Thus, a radius is easily estimated as
\begin{align}
    R_c \equiv b \sqrt{\frac{N_c}{4 \phi_{cp}}}.
\end{align}
The value $\langle R_c \rangle$ used in the main text is obtained by averaging over clusters in a configuration.
    
\clearpage
\pagebreak

\section{Equations of Motion}
    \subsection{\label{secZeroConcSI} Pair interactions (Zero Concentration)}

    \subsubsection{The dynamical system of a pair of force-aligning active particles}

\begin{figure}[h]
        \centering        \includegraphics[width=0.4\linewidth]{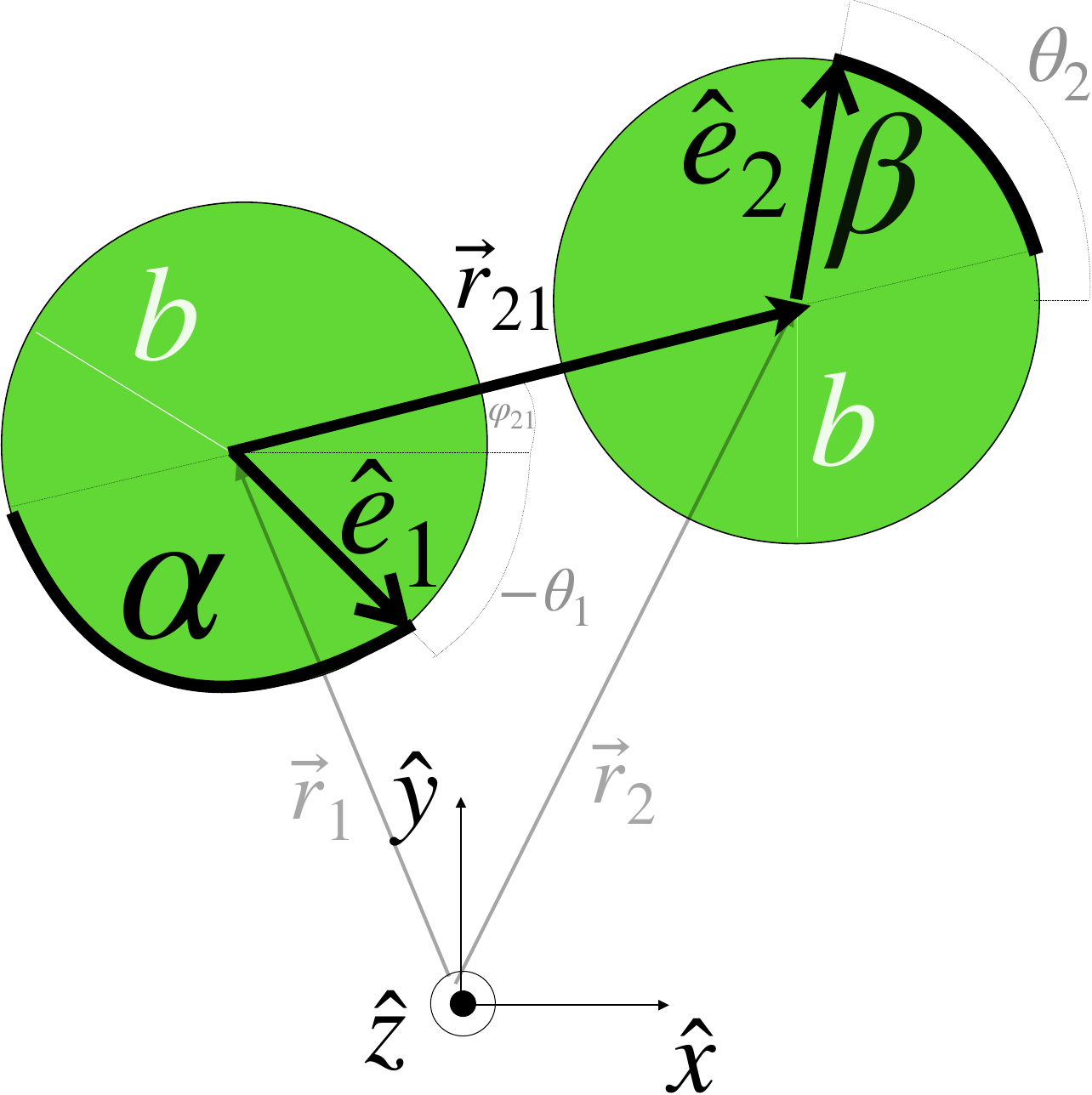}
        \caption{\footnotesize{Quantities of an interacting pair of force-aligning particles.}}
    \label{figPairSI}
    \end{figure}

    To derive the expected dynamics of two interacting force-aligning self-propelled particles we start by defining their dynamical quantities (see Fig.~\ref{figPairSI}). Particles 1 and 2 are located relative to the lab frame in positions $\vec{r}_1$, $\vec{r}_2$ respectively, with headings $\hat{e}_1$ $\hat{e}_2$ respectively. The headings are unit vectors which at the absence of an external force, set the self-propulsion at their nominal speed ($v_0 \hat{e}_i$). The headings are defined by the angle relative to the $\hat x$ axis, $\theta_1$ and $\theta_2$ where we follow the conventional right handed system (counter clockwise increase, see Fig.~\ref{figPairSI}). The center-to-center vectors are defined as the difference between the particles' positions:
    
    \begin{align}
        \vec{r}_{12} &\equiv \vec{r}_1 - \vec{r}_2=r\hat{r}_{12}\\
        \vec{r}_{21} &\equiv \vec{r}_2 - \vec{r}_1 = r\hat{r}_{21},\\
    \end{align}

    where the magnitude of the center-to-center separation is defined as 

    \begin{align}
        r \equiv |\vec{r}_{12}| = |\vec{r}_{21}|,\\
    \end{align}

    which define the unit vectors

    \begin{align}
        \hat{r}_{12} &\equiv \vec{r}_{12}/r\\
        \hat{r}_{21} &\equiv \vec{r}_{21}/r = -\hat{r}_{12}.
    \end{align}

    Each particle is subjected to the external force field acting by the other. In the case of the two identical particles, the forces are central and reciprocal (equal and opposite). The force on particle 1: $\vec{F}_{12} =v_0\Gamma/\mu \left(r_{12}\right)\vec{r}_{12} = -\vec{F}_{21}$. The profile function $\Gamma \left(r\right)$ is generally decaying ($\Gamma ' <0$) and non-negative $\Gamma \ge 0$, and makes an implicit function that defines the profile of the repulsive force field. It is chosen such that at the equilibrium separation (when the two self-propelled particles are pushing against each other head on), it is equal to unity: $\Gamma \left(2b\right) =1 $. In most simulations we used soft core (see Sec.~\ref{secSimulationSI}, and in the phase portrait we used a decaying exponent. $\Gamma$ is left implicit making the argument hold in general. 
    
    The velocities of the particles are then given by:
     \begin{align}
        \vec{v}_1 &\equiv \dot{\vec{r}}_{1} = v_0\left[\hat{e}_1 + \Gamma\left(r_{12}\right)\hat{r}_{12} \right] \\
        \vec{v}_2 &\equiv \dot{\vec{r}}_{2} = v_0\left[\hat{e}_2 + \Gamma\left(r_{21}\right)\hat{r}_{21} \right].
        \label{eqVelComponents}
    \end{align}
    
    Expressed in the relative vectors basis the velocities are:
    
    \begin{align}
        \dot{\vec{r}}_{12} &= \dot{r} \hat{r}_{12} +r\dot{\varphi} \hat{\varphi}_{12}\\
        \dot{\vec{r}}_{21} &= \dot{r} \hat{r}_{21} +r\dot{\varphi} \hat{\varphi}_{21}.
    \end{align}

    We next define the angles $\alpha$ and $\beta$, between the headings $\hat{e}_1$ and $\hat{e}_2$, and the center-to center vectors $\hat{r}_{12}$ and $\hat{r}_{21}$ respectively (see Fig.~\ref{figPairSI}). The connection between the heading angles, relative angles, and the center-to-center angles are:

    \begin{align}
       \theta_1 &= \alpha + \varphi_{12}\\
       \theta_2 &= \beta + \varphi_{21},
       \label{eqAnglesSI}
    \end{align}
    where the two global rotation angles are also connected by: $\varphi_{12} = \varphi_{21} +\pi$. This allows us to express the headings in the center-to-center coordinates by:
\begin{align}
       \hat{e}_1 &= \cos \alpha \hat{r}_{12} + \sin \alpha \hat{\varphi}_{12}\\
       \hat{e}_2 &= \cos \beta \hat{r}_{21} + \sin \beta \hat{\varphi}_{21}.
    \end{align}

    The two particles' headings are dynamically coupled to the velocities, which can be written as:

    \begin{align}
       \dot{\hat{e}}_1 &= \kappa \hat{e}_1\times \vec{v}_1 \times \hat{e}_1 \\
       \dot{\hat{e}}_2 &= \kappa \hat{e}_1\times \vec{v}_2 \times \hat{e}_2,
    \end{align}
    where the dynamics of the unit vectors in the lab frame coordinates are:
    \begin{align}
       \dot{\hat{e}}_1 &= \dot \theta_1 \hat{e}_1^\perp \\
       \dot{\hat{e}}_2 &= \dot \theta_2 \hat{e}_2^\perp.
    \end{align}

    The perpendicular vectors are also defined using the right-hand convention to satisfy:

    \begin{equation}
       \hat{e}\times\hat{e}^\perp = \hat{z},
    \end{equation}
    as they are restricted to the plane.

    After some simplification the relative velocity between the two particles can be shown to take the following form:

    \begin{equation}
       \dot{\vec{r}}_{12} = v_0\{\left[\cos \alpha +\cos \beta +2\Gamma\left(r\right) \right]\hat{r}_{12} + \left(\sin \alpha +\sin \beta\right)\hat{\varphi}_{12}\}.
    \end{equation}

    Which gives the dynamics of the two components (using Eqs.~\ref{eqVelComponents}):    

    \begin{align}
       \dot r  &= v_0\left[\cos \alpha +\cos \beta +2\Gamma\left(r\right) \right]\\
       \dot \varphi_{12} &= \frac{v_0}{r}\left[\sin \alpha +\sin \beta \right],
       \label{eqPairRPhiSI}
    \end{align}

    reproducing Eq.6 in the main text.

    To find the equations of the relative angles (i.e. $\dot \alpha$) we start plugging in the expressions of the velocities to the dynamics of their respective headings (i.e. $\vec{v}_1$ into equation of $\dot \hat{e}_1$). The dynamics of the headings' angles that follow are:
    
\begin{align}
       \dot \theta_1   &= -v_0\kappa \Gamma\left(r\right) \sin \alpha\\
       \dot \theta_2   &= -v_0\kappa \Gamma\left(r\right) \sin \beta.
    \end{align}

    Using the above expression along with the dynamics of the orientation of the center-to-center vector (Eq.~\ref{eqPairRPhiSI}), the geometric relations between the various angles (Eq.~\ref{eqAnglesSI}), and following some simplifications we arrive at:

    \begin{align}
     \dot\alpha &= -v_0 \left[\left(\kappa  \Gamma \left(r\right) + \frac{1}{r} \right) \sin \alpha + \frac{1}{r}\sin \beta\right]\\    
     \dot\beta &= -v_0 \left[\left(\kappa  \Gamma \left(r\right) + \frac{1}{r} \right) \sin \beta + \frac{1}{r}\sin \alpha\right],
    \end{align}

    which are Eqs.7-8 in the main text.

    \subsubsection{Fixed points and linear stability analysis}

    The dynamical equations describing the size of the center-to-center separation vector ($r$), and the angles ($\alpha$, $\beta$) of the headings relative to it (Eqs.6-8 in the main) have a few fixed points. Here we restrict our discussion to the fixed point where the two particles are facing each other ($\alpha = \beta = \pi$), and their centers-to-center separation is at a mechanical equilibrium $r = 2b$. It can be seen that Eqs.6-8 (main text) all vanish:

    \begin{equation}
        \dot r = \dot \alpha = \dot \beta =_{r=2b,\; \alpha=\beta=\pi}  0.
    \end{equation}

    Linearizing around this fixed point, gives the following linear system for the perturbation $(\delta r, \delta \alpha, \delta \beta)$,

    \begin{equation}
        \dot{\vec{x}}=
        \begin{bmatrix}
        \dot{\delta r}\\
        \dot{\delta\alpha}\\
        \dot{\delta \beta}
        \end{bmatrix}
        =
        \begin{bmatrix}
        2\Gamma'\left(2b\right) & 0 & 0\\
        0 & \kappa+\frac{1}{2b} & \frac{1}{2b}\\
        0 & \frac{1}{2b} & \kappa + \frac{1}{2b}
        \end{bmatrix}
        \begin{bmatrix}
        \delta r\\
        \delta\alpha\\
        \delta\beta
        \end{bmatrix}
        = A \vec{x}.
        \label{linear_system}
    \end{equation}

    Computing the characteristic equation of the linear system ($|A-\lambda I| =0$), yields the following eigenvalues:
    \begin{align}
        \lambda_1 &= 2\Gamma '\left(2b\right)\\
        \lambda_2 &= \kappa\\
        \lambda_3 &= \kappa + \frac{1}{b},
    \end{align}

    A stable fixed point is found when all the eigenvalues are negative. For a strictly repulsive pair, the force is decaying and therefor ($\lambda_1=\Gamma'<0$), and since the particles are convex circles ($1/b>0$), both $\lambda_2$ and $\lambda_3$ are negative when

    \begin{equation}
        \kappa + \frac{1}{b} < 0.
    \end{equation}

    This is the condition given in Eq.1 in the main text for the geometric criterion for effective attraction of a pair of force-aligning active particles.

    \subsubsection{Phase portraits}
        Phase portraits were drawn using Python matplotlib library with integration in both directions. For the plots, the chosen force profile of the particles was:

        \begin{equation}
            \Gamma_{\rm{portraits}} = e^{1-r/b}.
        \label{eqGammaPortait}
        \end{equation}
    
    \subsubsection{Bifurcation analysis}
We derive the normal form equation in terms of $\kappa b$ to make a local approximation of the system near the geometric condition. We adjust the Jacobian (Eq.~\ref{linear_system}) to make the bifurcation parameter $\kappa b$ explicit:

\begin{align}
    2b * A =
    \begin{bmatrix}
        4b\Gamma'(2b) & 0 & 0 \\
        0 & 2\kappa b+1 & 1 \\
        0 & 1 & 2\kappa b + 1
    \end{bmatrix}.
    \label{jacobian}
\end{align}

By solving $2b|A-\lambda I|=0$, we obtain
$\lambda_1 = 4b\Gamma'(2b)$, $\lambda_2 = 2\kappa b$, $\lambda_3 = 2(\kappa b+1)$. With corresponding eigenvectors (Fig. 3b in the main text):
\begin{align}
v_1 = 
    \begin{bmatrix}
        1  \\
        0 \\
        0 
    \end{bmatrix},
v_2 = 
    \begin{bmatrix}
        0 \\
        1  \\
         -1 
    \end{bmatrix},
v_3 = 
    \begin{bmatrix}
        0 \\
        1 \\
         1
    \end{bmatrix}.
\end{align}

When $\kappa b<-1$, all eigenvalues are real and negative, and the fixed point is a \textit{stable node}. At the critical value related to the geometric condition, $\kappa b=-1$, the system becomes unstable along $v_3$. This change of stability is indicative of a bifurcation.

We analyze the bifurcation as the parameter $\kappa b$ varies. We denote $2 \kappa b$ as $\gamma$, with the critical value $\gamma_c = -2$.

From Eq~\ref{jacobian} , we recognize that in the linear approximation, $\dot{r}$ is decoupled and independent of the bifurcation parameter. Therefore, we focus on $\dot{\alpha}$, and $\dot{\beta}$ follows similarly. We start from Eq. 7 from the main text:

$$\dot{\alpha}=-v_0[(\kappa\Gamma(r)+\frac{1}{r})\sin\alpha+\frac{1}{r}\sin\beta]$$

We rewrite this in terms of the bifurcation parameter $\gamma$:
\begin{align}
    \dot{\alpha}=-\frac{v_0}{r}[
    (\gamma+1)\sin\alpha + \sin\beta],
\end{align}

And reduce it to the dimensionless form: 
\begin{align}
    \dot{\alpha}=-[
    (\gamma+1)\sin\alpha + \sin\beta],
\end{align}

To analyze the bifurcation, we Taylor expand the vector field of the governing equation $V(\alpha, \gamma)$ around the critical value $\Tilde{\gamma}=\gamma + 2$ and the fixed point. Truncating higher-order terms and simplifying results in:

\begin{align}
    \dot{\alpha}= \alpha(\Tilde{\gamma}  + \frac{\alpha^2}{6}).
    \label{normal_form}
\end{align}

As formulated in \cite{Crawford1991}. This corresponds to the canonical form of a \textit{pitchfork bifurcation}, where: $\frac{\delta^2 V}{\delta \alpha \delta \gamma}(0,0) \neq 0$ and $\frac{\delta^3 V}{\delta \alpha^3}(0,0) \neq 0$:
\begin{align}
    \dot{\alpha}=\frac{\delta^2 V}{\delta \alpha \delta \gamma}(0,0)\Tilde{\gamma} \alpha [1 + O (\gamma, \alpha)] +
    \frac{\delta^3 V}{\delta \alpha^3}(0,0)\frac{\alpha^3}{3!} [1 + O (\gamma, \alpha)].
\end{align}

At the critical value, the fixed point changes stability, and two symmetrical fixed points with the opposite stability from the changing fixed point are created.

The change in stability can be described as a sub-critical pitchfork bifurcation. This is derived from the sign of $\epsilon_1$ and $\epsilon_2$ from the normal form (Eq.~\ref{normal_form}):
$$\dot{\alpha}=\alpha(\epsilon_1\Tilde{\gamma}+\epsilon_2 \alpha^2).$$

$\epsilon_1$ and $\epsilon_2$ denote the sign of $\frac{\delta^2 V}{\delta \alpha \delta \gamma}(0,0)$ and $\frac{\delta^3 V}{\delta \alpha^3}(0,0)$ respectively \cite{Crawford1991}.
From (Eq.~\ref{normal_form}), $\epsilon_1 = \epsilon_2 = 1$, describing a \textit{subcritical pitchfork bifurcation}.
Therefore, when $\kappa b$ is lowered and the critical value $\kappa b = -1$ is reached, the fixed point becomes stable, and two unstable fixed points arise.

To get the bifurcation diagram in Figure 3a in the main text, we derive the unstable fixed points. These are only correct near the bifurcation as it is a local approximation. By filling in $\Tilde{\gamma}=\gamma+2$, $\gamma = 2\kappa b$, and setting the normal form (Eq.~\ref{normal_form}) to 0, we get the following solutions:  
 \begin{align}
     \alpha = \pm\sqrt{-12(\kappa b+1)}.
 \end{align}

It is important to note that the extra fixed points cancel each other out when considering the symmetry between $\dot{\alpha}$ and $\dot{\beta}$.
    
\section{\label{secFiniteConcentration} Many-Body Dynamics (Finite Concentration)}
\subsection{Mass Conservation}
At a finite concentration, a self-propelled particle with curvity $\kappa$ becomes effectively attracted to a cluster of radius $a$, when the geometric condition for the attraction (Eq.1 in the main text) is satisfied for that cluster size:
\begin{equation}
    \kappa + 1/a < 0.
    \label{eqAttractionToCluster}
\end{equation}
Assuming that when there are no individual free particles in the system (singlet fraction $\rightarrow 0$) the system coarsens into clusters of similar size ($a$), that are separated by similar distance $L$, mass conservation connects the average filling fraction, $\Phi$ to the cluster size by:

\begin{equation}
    \Phi \approx \frac{a^2}{L^2}.
    \label{eqMassConservation}
\end{equation}

The basin of attraction (Fig.2 in the main text), shows that the effective attraction to the target stretches by some factor, $A$ beyond the target size. We shall call it the capture distance and assume that this is the same distance that sets the steady state separation of the clusters: $L = Ab$. we see that there is a finite range of distances from the target's center that effective attraction of the target (cluster) extends beyond the equilibrium distance. Generally, this factor, $A$, depends on the specific shape of the potential repulsion.

Combining Eqs.~\ref{eqAttractionToCluster} and \ref{eqMassConservation} we arrive at the square root criterion of the phase boundary at finite filling fraction presented in the text:

\begin{equation}
    Ab\kappa \sqrt{\phi} < -1,
    \label{eqFinitConcentrationSI}
\end{equation}

(Eq.12 in the main text).

\subsection{Kinetic theory\label{sec:kinetic}}

Another approach to find Eq.~\ref{eqFinitConcentrationSI} is to use a kinetic argument.
Assume that a near-critical cluster with radius $a = R_c^{-} = 1/|\kappa|$ is present in the system.
At the simplest possible level of approximation, a particle living on its outer rim can escape the cluster after a time that scales roughly like the perimeter of the cluster divided by the self-propulsion speed
\begin{align}
    \tau_{out} \sim \frac{2 \pi R_c(\phi)}{v_0}.
\end{align}
On the other hand, a new particle will hit the cluster after a time given by the mean typical distance between 2 particles at the chosen density, also divided by the self-propulsion speed,
\begin{align}
    \tau_{in} \sim \frac{1}{\rho^{1/d} v_0} = \frac{b}{\phi^{1/d} v_0}.
\end{align}
Equating the two times gives a kinetic criterion for a cluster to be stable,
\begin{align}
    \frac{2 \pi R_c(\phi_c)}{v_0} \sim \frac{b}{\phi_c^{1/d} v_0}
\end{align}
or, after a few simplifications, using the link between $R_c$ and $\kappa_c$, and removing unitless prefactors
\begin{align}
    \kappa_c \sim \sqrt{\phi_c}.
\end{align}

A more involved approximation requires to use the dynamics to estimate the escape time.
Consider the full equations of motion (in reduced polar variables) for a particle and a cluster, treated as a fixed obstacle,
\begin{align}
    \dot{r} &= v_0 (\cos \psi + \Gamma(r)) \\
    \dot{\psi} &= - v_0 (\kappa \Gamma(r) + \frac{1}{r})\sin\psi.
\end{align}
For the purpose of this estimate, assume that we focus on the following situation: the particle is at the edge of the cluster, at $r = R_c$, with a subcritical cluster $R_c \leq -1/ \kappa$, feeling a constant force $\Gamma_0$, and initially at a reduced angle almost pointing into the cluster, $\psi = \pi - \varepsilon$.
We seek, as an escape time, the time it takes the particle to reach $\psi = \pi/2$, so that it can slide against the cluster.
Then, the only differential equation to consider is
\begin{align}
     \dot{\psi} = - v_0 (\kappa \Gamma_0 + \frac{1}{R_c})\sin\psi, \label{eq:kinetic_angle}
\end{align}
solved with the initial condition
\begin{align}
    \psi(t=0) = \psi_0 = \pi - \varepsilon.
\end{align}
The solution may be obtained analytically as
\begin{align}
    \psi(t) = 2 \arctan\left[ e^{- (\kappa \Gamma_0   + \frac{1}{R_c}) v_0 t } \tan \frac{\psi_0}{2}\right].
\end{align}
Then, solving for $\psi(\tau_{out}) \equiv \pi/2$ yields
\begin{align}
    \tau_{out}(\psi_0) = \frac{R_c}{v_0}\frac{\ln \tan (\psi_0/2)}{1 + \kappa \Gamma_0 R_c}.
\end{align}
As expected from the fact that there is a fixed point at $\psi_0 = \pi$ and $\kappa R_c = 1$ that is attractive along the $\psi_0 = \pi$ line, this value diverges as $\psi_0 \to \pi$.
However, for any finite $\varepsilon$, it is a finite time that may yield interesting scalings.
Note in particular that the scaling of this escape time with respect to $1 + \Gamma_0 \kappa R_c$ matches well the kissing time results from pair simulations, Fig.~\ref{fig:kissing}. 

Equating this $\tau_{out}$ to $\tau_{in}$, we get an expression for a critical $\kappa_c$ separating two different regimes of ordering of the events ``the particle detaches from the cluster'' and ``another particle hits''
\begin{align}
    \kappa_c b = \frac{-1 + \sqrt{\phi_c} R_c/b \ln \tan(\psi_0 / 2) }{\Gamma_0 R_c/b}.
\end{align}
In particular, assuming that the particle started from the region where it could compensate self-propulsion by repulsion, $\Gamma_0 = 1$, and that we look at the onset of clustering from a single particle, $R_c = b$, this expression yields
\begin{align}
    \kappa_c b = -1 + \sqrt{\phi_c} \ln \tan(\psi_0 / 2).
\end{align}
Therefore, on average over collisions between particles, the critical line can be well approximated by
\begin{align}
    \kappa_c b = -1 + C \sqrt{\phi_c},
\end{align}
with $C$ a constant (Eq.12 in the main text).
This is consistent with the observed shape at zero noise and small densities, where in practice the full line separating domains is well approximated by $\kappa_c b + 1 \propto \sqrt{\tanh C \phi_c}$.

Note that a way to get a (wrong but with the right order of magnitude) estimate of the prefactor is to estimate an average escape time using
\begin{align}
    \langle \ln \tan(\psi_0 / 2) \rangle \sim \int_{\pi/2}^{\pi} d\psi \frac{2}{\pi} \ln \tan(\psi_0 / 2) = 4 G / \pi 
\end{align}
with $G \approx 0.915\ldots$ Catalan's constant.

\subsection{Curvature of free trajectories}

Consider a free particle following the dynamics prescribed by (Eqs.9,10 main text) in the absence of an external force, but in the presence of noise.
In dimensional units,
\begin{align}
    \frac{d}{dt}\vec{r}_{i} &= v_0 \hat{e}_i\label{eqSingleFreeForce}\\
    \frac{d}{dt} \hat{e}_i &= \kappa  \, \hat{e}_i \times \left(\vec{v}_i\times\hat{e}_i\right) + \sqrt{2 D_r}\xi_i(t) \hat{e}_i^{\perp}.
    \label{eqSingleFreeAlign}
\end{align}
To evaluate the effect of noise on the curvature of the trajectory, assume $\kappa \to 0$, which yields the well-known Active Brownian Particle dynamics,
\begin{align}
    \frac{d}{dt}\vec{r}_{i} &= v_0 \hat{e}_i\label{ABPForce}\\
    \frac{d}{dt} \hat{e}_i &= \sqrt{2 D_r}\xi_i(t) \hat{e}_i^{\perp}.
    \label{ABPAlign}
\end{align}
The curvature of this trajectory is defined via
\begin{align}
    K(s) \equiv \left|\frac{d^2 \bm{r}}{ds^2} \right|,
\end{align}
with $s = v_0 t$ the curvilinear coordinate.
Using the definition of the dynamics yields
\begin{align}
    K(s) = \frac{1}{v_0} |\dot{\theta}(s)|,
\end{align}
a stochastic quantity.
To evaluate the typical value of this quantity, it is convenient to introduce an RMS curvature,
\begin{align}
    \overline{K} \equiv \frac{1}{v_0}\sqrt{\langle (\theta(\tau) - \theta(0))^2/\tau^2\rangle}
\end{align}
with $\tau$ some typical microscopic time scale.
Using the definition of angular dynamics, $(\theta(\tau) - \theta(0))^2 = 2 D_r \tau$.
Choosing as a microscopic time scale $\tau \equiv d / v_0$ the time it takes a particle to move by some distance $d$, this expression reduces to
\begin{align}
    \overline{K} = \sqrt{\frac{2 D_r }{ d v_0}}
\end{align}
Introducing the persistence length $l_p = v_0 / D_r$ as well as the same definition of the Péclet number as in the main text, $Pe = v_0 / (\sigma D_r)$ with $\sigma$ the repulsive range, this can be rewritten as
\begin{align}
     \overline{K} = \sqrt{\frac{2}{ d l_P}} = \sqrt{\frac{2}{ d \sigma Pe}}.
\end{align}
The prefactor from the main text in the scaling of $\kappa_c$ can be recovered exactly by equating
\begin{align}
    \frac{2}{b\sqrt{Pe}} = \sqrt{\frac{2}{ d \sigma Pe}},
\end{align}
yielding a choice for $d$, $d = b^2 / (2\sigma)$ which, for our choice of repulsion, is well approximated by $d \approx b/4$.

\section{Video captions}

We here provide captions for the various supplementary videos attached to this paper.
All videos were obtained using the numerical method of Sec.~\ref{secSimulationSI} and saving regularly-spaced snapshots of the dynamics.
All videos were obtained at $\phi = 0.2$, $N = 8192$, and using time intervals $v_0 \Delta t /\sigma = 10$ between frames -- a time scale larger than that of self-propulsion but smaller than that of cluster assembly and flock motion.
In all videos, the color of particles encodes the orientation of its self-propulsion vector, as per the color wheel used in the snapshots in the main text.

\begin{itemize}

    \item \textit{Movie S1 - NoisySmallClusters.mp4} -- Dynamics at $Pe = 100$, and $\kappa b = -0.4$. In the presence of noise, fast clustering occurs, yet a background fluid persists and continuously exchanges particles with the cluster. For this value of $\kappa$, multiple clusters form in steady state.

    \item \textit{Movie S2 - NoisySmallClusters-Zoomin.mp4} -- Zoomed-in detail of the previous video, showing the dynamics of a single small cluster up close. For this video, we render particles as $3d$ colored dielectric media using the same ray-tracing engine as in Ref.~\cite{Casiulis2022a}.

     \item \textit{Movie S3 - MultipleClusters\_ZeroNoise.mp4} -- Dynamics at zero noise, $1/Pe = 0$, and $\kappa b = -0.4$. In the absence of noise, fast clustering occurs and empties the fluid domain, as often in persistent self-propelled particle systems, see \textit{e.g.} Ref.~\cite{Casiulis2021}. For this value of $\kappa$, multiple clusters form in steady state.
    
    \item \textit{Movie S4 - NoisySingleCluster.mp4} -- Dynamics at $Pe = 100$, and $\kappa b = -0.2$. In the presence of noise, fast clustering occurs, yet a background fluid persists and continuously exchanges particles with the cluster. For this value of $\kappa$, a single large cluster forms in steady state.

    \item \textit{Movie S5 - SingleCluster\_ZeroNoise.mp4} -- Dynamics at zero noise, $1/Pe = 0$, and $\kappa b = -0.2$. In the absence of noise, fast clustering occurs and empties the fluid domain, as often in persistent self-propelled particle systems, see \textit{e.g.} Ref.~\cite{Casiulis2021}. For this value of $\kappa$, a single cluster forms in steady state.
    
    \item  \textit{Movie S6 - Flocks\_ZeroNoise.mp4} -- Dynamics at zero noise, $1/Pe = 0$, and $\kappa b = 0.2$. In the absence of noise, flocking quickly sets in with a near-unit average magnetization of self-propulsion orientations.

    \item  \textit{Movie S7 - NoisyFlock.mp4} -- Dynamics at moderate noise, $Pe = 100$, and $\kappa b = 0.2$. Due to finite noise, flocking is suppressed so that the total magnetization is non-zero yet not 1.

    \item \textit{Movie S8 - ClusterFlockTransition.mp4} -- Video corresponding to the dynamics described in Fig.~5 of the main text. Noise is set to $Pe = 50$, and $\kappa$ is switched between the values $\kappa b \pm 0.09$ at times $v_0 T_n / \sigma = 1000 n$, with $n = 1, 2$.

    \item \textit{Movie S9 - ClusterFlockTransition-Zoomin.mp4} -- Zoomed-in detail of the previous video, showing the formation of a cluster and its transition into a flock. For this video, we render particles as $3d$ colored dielectric media using the same ray-tracing engine as in Ref.~\cite{Casiulis2022a}.

\end{itemize}

\bibliography{clusterFlock, supp}